\documentclass[5p,twocolumn]{elsarticle}
\usepackage{multicol}
\usepackage{graphicx}
\usepackage{subcaption}
\usepackage{amssymb}

\journal{Powder Technology}

\begin{document}

\begin{frontmatter}

\title{Measuring segregation characteristics of industrially relevant granular mixtures: Part I -- A continuum model approach}
\author[meche]{Alexander M. Fry}
\author[pandg]{Vidya Vidyapati}
\author[pandg]{John P. Hecht}
\author[meche]{Paul B. Umbanhowar\corref{cor1}}
\author[meche,cheme,nico]{Julio M. Ottino}
\author[meche,cheme,nico]{Richard M. Lueptow}
\cortext[cor1]{umbanhowar@northwestern.edu}
\address[meche]{Department of Mechanical Engineering, Northwestern University, Evanston, IL 60208, USA}
\address[pandg]{Global Engineering, Procter \& Gamble Company, West Chester, OH 45069, USA}
\address[cheme]{Department of Chemical and Biological Engineering, Northwestern University, Evanston, IL 60208, USA}
\address[nico]{Northwestern Institute on Complex Systems (NICO), Northwestern University, Evanston, IL 60208, USA}
\date{\today}

\address{}

%

\begin{abstract}
We present a method to estimate the segregation parameter, $S,$ a key input in a continuum transport model of particulate flows.  $S$ is determined by minimizing the difference between measured and model-predicted concentration profiles. To validate the approach, we conduct discrete element method simulations of size-bidisperse mixtures in quasi-2D bounded heap flow; the resulting data show that $S$ calculated from concentration profiles is consistent with the directly measured value.  The method's accuracy depends critically on the velocity profile during filling, but only weakly on the diffusion coefficient. When the velocity profile is nominally spanwise invariant, the error between estimated and measured $S$ is $10\%$. This method is intended for practical application (described in Part II), so we restrict characterization of the velocity profile to that which can be readily determined experimentally, and explore the sensitivity of concentration profiles to variation of the gap between the sidewalls of the heap.  
\end{abstract}


\end{frontmatter}

\section{Introduction\label{Introduction}}
Particles in dense flowing mixtures tend to rearrange, or segregate, often resulting in spatial inhomogeneities if the particle species vary in size, density, shape, or other properties~\cite{ottinoKhakhar2000,gray2018,meier2007,umbanhowar2019}. Recent developments~\cite{fan2014,hillTan2014,bertuola2016} in transport equation continuum models~\cite{bridgwater1985,dolgunin1995,grayThornton2005} for particle size or density segregation make it possible to accurately predict segregation of granular materials in flow geometries that are applicable to industrial processes. In these continuum models, the degree of segregation results from a competition between advection (mean flow), random collisional diffusion, and species-specific segregation within the flow. The flow kinematics can be obtained via discrete element method (DEM) simulations~\cite{fan2013,fan2014,schlick2015}, based on theory~\cite{schlick2015b}, or measured experimentally~\cite{lueptow2000,komatsu2001,jesuthasan2006,eckart2003,wiederseiner2011}. The segregation and diffusion parameters, however, are more difficult to determine, and, to date for heaps, have only been obtained from DEM simulations~\cite{schlick2015,xiao2016,zhao2017,jones2018}. The segregation coefficient scales with the logarithm of the diameter ratio between spherical particles~\cite{schlick2015}, the length ratio of equal diameter rod-like particles~\cite{zhao2017}, or the density ratio between equal diameter spherical particles~\cite{xiao2016}. The diffusion coefficient scales with the product of the local shear rate and the square of the mean particle diameter for both monodisperse particles~\cite{bridgwater1980,hsiau1999,utter2004} and bidisperse particle mixtures~\cite{fan2013,schlick2015,fan2015,xiao2016}. 

While DEM simulations have provided reasonable estimates of segregation and diffusion parameters for bidispserse mixtures of idealized spherical particles for which simulation parameters are known or can be estimated (such as for mm-sized glass particles), many practical mixtures are comprised of non-spherical particles with a variety of different and unknown physical properties, e.g., stiffness, friction coefficient, and density. Furthermore, the segregation parameters for mixtures of realistic particles varying simultaneously in size, density, and shape are difficult to obtain via DEM simulations. The objective of this work is to bring the power of the continuum segregation model to industrially relevant granular mixtures. Rather than assuming the physical properties of each particle species in the segregating mixture and then using DEM simulation of representative particles to estimate the model parameters, we demonstrate a method to determine the segregation coefficient for real particle mixtures from experiment. The segregation coefficient determined from a single experiment can then be used to model segregation in various industrial flow scenarios.

Although techniques exist for measuring particle velocity optically at the transparent sidewalls or surface of a granular flow~\cite{lueptow2000,komatsu2001,jesuthasan2006,eckart2003} (which makes it possible to determine the bulk particle flow), it is generally quite challenging to characterize individual grain-scale rearrangements within the flowing layer precisely enough to obtain segregation or diffusion data. Thus, we explore an alternative method.  In this approach, a small-scale quasi-2D bounded heap experiment is performed with the bidisperse mixture of interest. Then, the segregation and diffusion parameters are found such that the spatial dependence of the species concentrations generated by the continuum model matches that for particles deposited in the bounded heap experiment. The background for this approach is described in Section~\ref{segregationTools}. In Section~\ref{backfitMethod}, we explore this approach by conducting DEM simulations of bounded heap formation similar to those studied experimentally in Part II of this work~\cite{fry2019b}. We then confirm the validity of the approach by comparing parameters estimated from the deposited concentration profiles using the continuum model (as would be done in a heap experiment) with parameters measured locally in the flowing layer, which can only be done using data from DEM simulations.

Lastly, in support of the experimental application of this technique in Part II of this study~\cite{fry2019b}, we explore the influence of system geometry on the efficacy of the technique in Section~\ref{flowConditions}. Specifically, we vary the gap between sidewalls and measure its impact on the spanwise variation of the velocity profile and resulting concentration profiles~\cite{jop2005,isner2017}, since it is usually impractical to implement techniques, such as PEPT~\cite{lim2003} or x-ray~\cite{zaman2016}, to measure the flow away from the sidewalls in industrially-useful experimental apparatus.

\section{Modeling segregation in granular flows}\label{segregationTools}
We utilize three established tools for studying segregation in granular flows: the quasi-2D bounded heap, DEM simulations, and a scalar transport equation based continuum model of segregation. The objective is to use DEM simulations of segregation in bounded heap flow to provide the kinematics and the concentration fields to the continuum-based segregation model, so we begin with a brief description of the continuum model.

\subsection{Continuum model for segregation}\label{continuumModelSection}
While experiments and DEM simulations can provide useful information about particle segregation, it is difficult to scale these techniques up to model large scale industrial processes. An alternative is to use continuum models to describe the segregation process. Previous research has shown that a continuum model approach can accurately predict segregation in flows of glass-like, frictional particles in multiple geometries, including bounded heaps~\cite{fan2014,schlick2015,xiao2016,zhao2017}, chutes~\cite{schlick2015c,schlick2016,deng2018}, cylindrical tumblers~\cite{schlick2016,deng2019}, and planar shear cells~\cite{fry2019}. 

The model is a scalar transport equation with an additional term that accounts for a species-dependent segregation relative to the mean flow~\cite{bridgwater1985}. Although many similar variants have been proposed~\cite{gray2018,umbanhowar2019}, we utilize the following form~\cite{fan2014}:
\begin{equation}\label{continuumModel}
\frac{\partial c_{i}}{\partial t} + \nabla \cdot (\mathbf{u} c_{i})+ \frac{\partial}{\partial z}(w_{s,i} c_{i})= \nabla \cdot (D\nabla c_{i}),
\end{equation}
where $\mathbf{u}$ is the velocity field and $c_i$ is the concentration of species $i$. In order to solve the model, one must provide the diffusion coefficient, $D$, and the inter-species percolation velocity, $w_{s,i}$, which accounts for the segregation, as well as the mean velocity profiles in the flowing layer, $u$ (mean flow in streamwise direction) and $w$ (mean flow normal to the heap surface).

\subsection{Quasi-2D bounded heap kinematics}\label{quasi2DBoundedHeap}

\begin{figure}[ht]
\begin{center}
\includegraphics[width=\columnwidth]{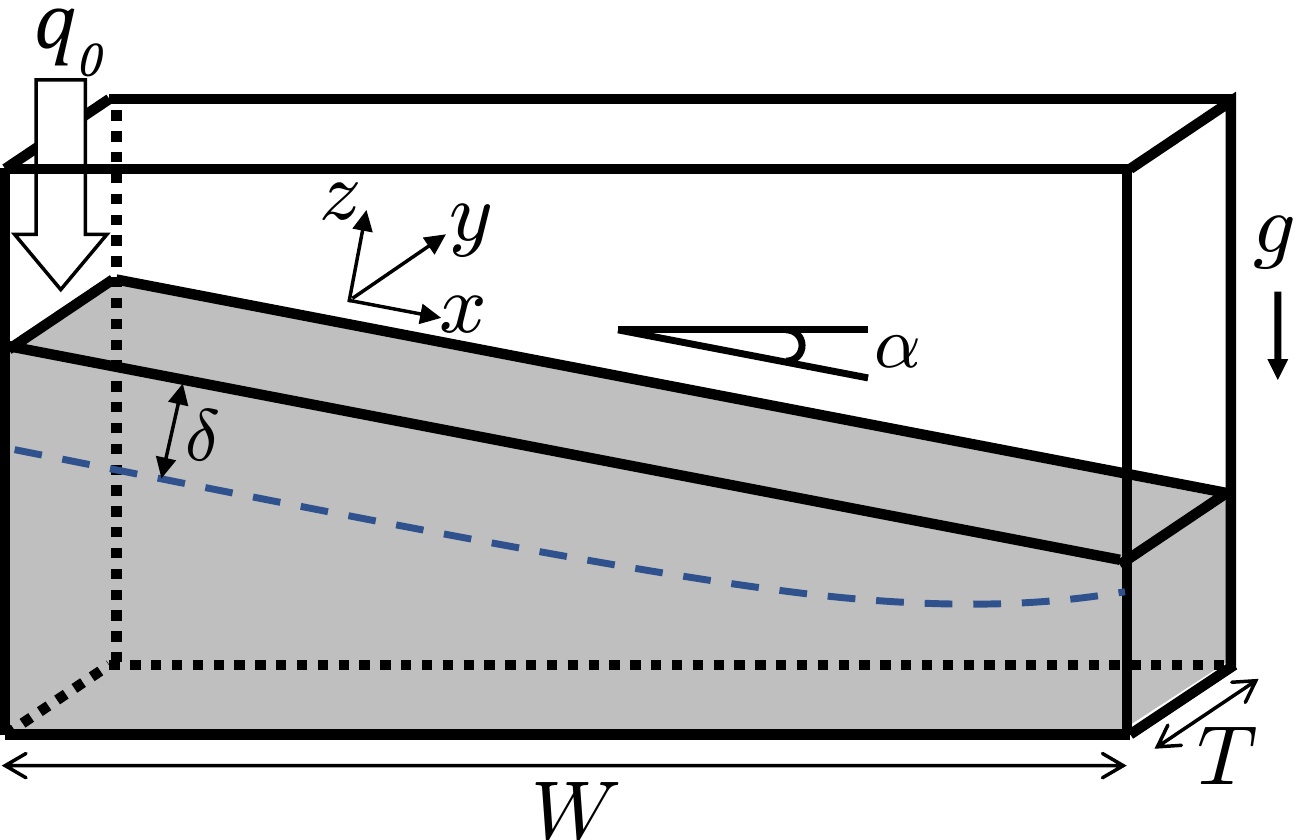}
\caption{Sketch of rectangular bounded heap geometry with length $W$ and spanwise gap width $T$. Particles are fed into one end of the bin at 2D bulk volumetric feed rate, $q_0$, to form a heap with angle of repose $\alpha$ relative to horizontal. Dashed curve represents the thickness $\delta$ of the flowing layer of particles at the surface of the growing heap. Below the dashed curve, particles are effectively static.}
\label{quasi2DBoundedHeapGeometryFigure}
\end{center}
\end{figure}

The technique we propose is based on flow in a quasi-2D bounded heap, which is well-characterized, and allows a wide range of flow conditions and relative particle concentrations. Concurrently, it is a close proxy for industrial flow settings like hopper/bin filling and chute flows. A detailed exposition on the bounded heap geometry and its use in granular materials research can be found in the review of Fan et al.~\cite{fan2017}. Briefly, the bounded heap is a rectangular bin into which particles are fed (at one end for a one-sided bounded heap, which is referred to here more generally as a bounded heap). The flow is easily observed through transparent sidewalls, which allows flow measurement during heap formation using standard optical techniques~\cite{lueptow2000,eckart2003,jesuthasan2006} and visual inspection of the segregation patterns in the deposited heap~\cite{fan2012,johanson2014}.

Figure~\ref{quasi2DBoundedHeapGeometryFigure} is a sketch of a typical rectangular bounded heap geometry. Particles are fed into the system at one end at a two-dimensional volumetric feed rate, $q_0=Q/T$, where $Q$ is the bulk volumetric feed rate and $T$ is the gap width between the sidewalls. In this study, all initial conditions are fully mixed, but the continuum model techniques have been shown to apply to bounded heap flow with initially segregated~\cite{fan2014} or time-periodic~\cite{lueptow2017} inlet conditions. After entering the system, particles flow down the heap away from the feed zone and deposit with angle $\alpha_{static}$ relative to the horizontal. During heap formation, newly added particles flow in a thin layer of thickness $\delta$ at an angle $\alpha_{dynamic}>\alpha_{static}$, on top of a largely static region of particles that rises over time at rise velocity $v_r=q_0/W$, where $W$ is the length of the bounding box. Defining the precise location of the bottom of the flowing layer is somewhat arbitrary, as the streamwise velocity typically decays exponentially with depth~\cite{fan2013,komatsu2001}. Here, as in previous studies~\cite{fan2013,fan2014,schlick2015,xiao2016}, the flowing layer thickness at streamwise location $x$ is defined as $\delta(x)=z_{surf}(x)-z(x,u=0.1u_{surf})$, that is, the distance below the free surface at which the velocity is 10\% of the surface velocity, $u_{surf}$, at that streamwise location. 

In this, Part I of our study, quasi-2D bounded heap flows of granular materials are simulated using a soft-sphere discrete element method (DEM) (see \ref{DEMSimulations}). A major benefit of DEM simulations is that the motion of particles can be measured at \textit{all locations} during heap formation. As a result, the segregation and diffusion parameters can be measured directly throughout the flowing layer of the bounded heap~\cite{fan2014,schlick2015,umbanhowar2019}.

The velocity profiles measured in the DEM quasi-2D bounded heap simulations here and in previous studies~\cite{fan2013,fan2014,schlick2015,xiao2016} are well-approximated by exponential profiles
\begin{equation}\label{flowingLayerKinematicsStreamwise}
u(x,z)= \frac{kq}{\delta(1-e^{-k})} (1-x/L)e^{kz/\delta}
\end{equation}
and
\begin{equation}\label{flowingLayerKinematicsNormal}
w(x,z)= \frac{q}{L(1-e^{-k})} (e^{kz/\delta}-1),
\end{equation}
where $k=2.3$ and $L$ is the streamwise length of the flowing layer, $L=W/ \cos{\alpha}$. While the profiles in this particular geometry are well characterized by exponential profiles, there exist scenarios in which other velocity profiles are possible, including linear~\cite{fan2013,gdrmidi2004,socie2005} and Bagnold~\cite{deng2018}. The velocity profile used in the continuum model must approximate that in the geometry being studied, and can usually be determined in geometries with transparent sidewalls using high speed imaging and image differencing techniques (e.g., Particle Tracking Velocimetry~\cite{jesuthasan2006} or Particle Image Velocimetry~\cite{lueptow2000,eckart2003}).

The local flowing layer thickness depends on the local flow rate, $q(x),$ as $\delta(x) \sim q(x)^{\beta}$~\cite{jop2005,fan2013} and is, consequently, a function of streamwise location~\cite{khakhar2001}, since $q$ decreases linearly with position along the heap as particles are uniformly deposited out of the flowing layer. The exponent $\beta$ in open heap flow (i.e.\ no end wall) is about $2/7\approx 0.286$ for channels between 20 and 600 particle diameters wide~\cite{jop2005}, while in a bounded heap flow (i.e.\ with an end wall) $\beta$ is about 0.22 for channels between 5 and 10 particle diameters wide~\cite{fan2013}. Here, $\beta=0.24$ for spanwise gap width, $T,$ greater than or equal to ten particle diameters, but $\beta$ decreases with decreasing $T$ for $T$ smaller than about ten particle diameters. For the simulations in Section~\ref{backfitMethod}, where $T$ is about five large particle diameters, $\beta=0.15$ best matches the data.

Since local flow rate decreases linearly with streamwise displacement in the quasi-2D heap, $\delta(x)=\delta_0 (x/L)^{\beta}$, where $\delta_0$ is the maximum flowing layer thickness in the upstream portion of the heap. While the upstream (maximum) flowing layer thickness defines our reference value $\delta_0$ in the relation, the local flow rate dependence relation works equally well for reference values chosen at other streamwise locations. 

We note that in some past work,  $\delta$ was assumed to be constant (i.e.\ $\beta=0$)~\cite{fan2014,schlick2015}. While this simplification is technically incorrect~\cite{fan2013}, the resulting continuum model predictions of segregation patterns match those observed in quasi-2D bounded heaps~\cite{fan2014,schlick2015} because $\beta$ in quasi-2D geometries is relatively small.  (In 3D heap flows, however, modeling streamwise $\delta$ variation is crucial as $\beta$ is considerably larger there, i.e.\ $\beta \approx 0.5$~\cite{isner2017,isner2019}.) Section~\ref{localVariation} compares the impact of constant vs.\ spatially varying diffusion and flowing layer thickness variations on the parameter estimation method proposed here. The predictions using locally-varying flowing layer thickness are slightly more accurate than the predictions using a constant flowing layer thickness, so locally varying models for $\delta$ and $D$ will hereinafter be used. However, a constant flowing layer thickness assumption provides relatively accurate results.

One final note on flowing layer kinematics in the quasi-2D bounded heap, which is important in Part II of this work~\cite{fry2019b}, concerns the use of either the flowing layer thickness, $\delta$, or the surface velocity, $u_{surf}$, to characterize the velocity profile (Eqns.~\ref{flowingLayerKinematicsStreamwise}, \ref{flowingLayerKinematicsNormal}). Previously work~\cite{fan2014,schlick2015,xiao2016} measured $\delta$ in the flowing layer of DEM simulations to provide the dimensional scale of the flowing layer velocity profile. However, mass conservation implies that once the characteristic form of the depthwise dependence of streamwise velocity is known (in the case of Eqns.~\ref{flowingLayerKinematicsStreamwise} and \ref{flowingLayerKinematicsNormal} it is exponential), either the flowing layer thickness, $\delta$, or the surface velocity, $u_{surf}$, can be used to set the dimensional scale of the velocity profile. Practically, this allows characterization of the velocity profile based on measurements of the streamwise velocity at the free surface of a bounded heap, rather than measuring the velocity profile at a transparent sidewal. In Part II of this work~\cite{fry2019b}, where we experimentally implement the parameter estimation method described here, the surface velocity measurement approach is used to determine the flowing layer thickness.  This eliminates problems with measurement at transparent sidewalls in an experimental apparatus, including boundary effects such as near wall deviation of the velocity profile from the profile in the bulk, and the influence of electrostatic-induced particle attraction to the sidewalls that can distort the velocity profile measurement (as is especially prevalent in flows with particle diameters less than $100 \, \mathrm{\mu m}$).

\subsection{Segregation and diffusion coefficients}\label{coefficients}

\begin{figure}[ht]
\begin{center}
\includegraphics[width=\columnwidth]{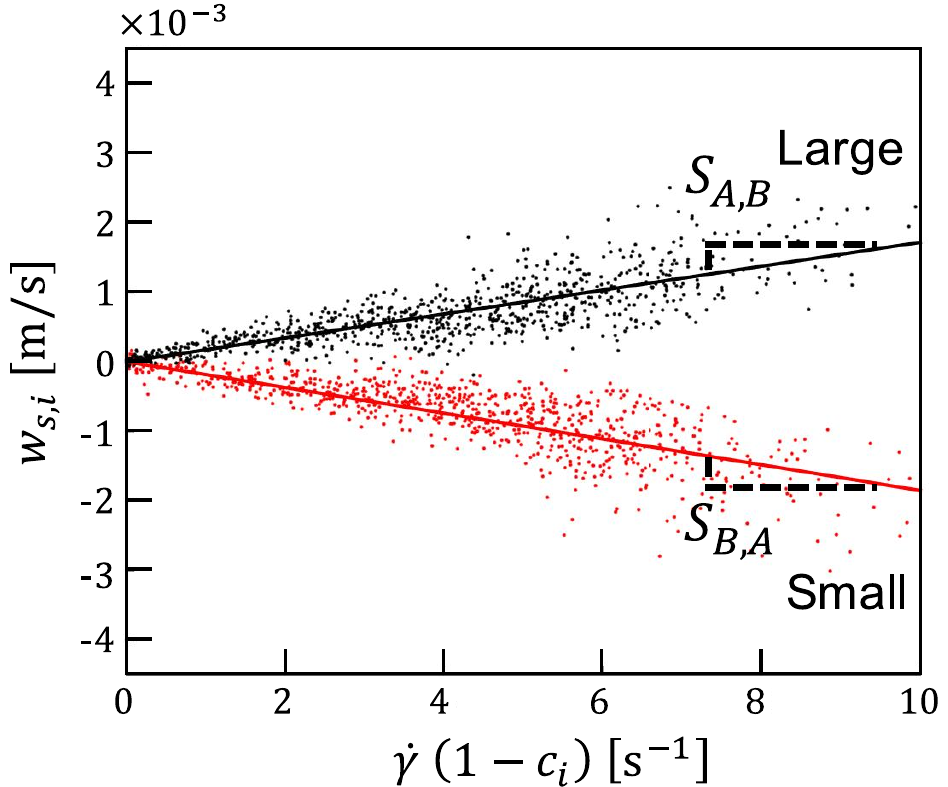}
\caption{Percolation velocity, $w_{s,i}$, (i.e., local free-surface-normal velocity of species $A$ (large) or $B$ (small) relative to the mean flow) vs.\ the product of shear rate and local species concentration $\dot{\gamma} (1-c_i)$. Data are measured in the flowing layer of a quasi-2D bounded heap DEM simulation of length $W= 60 \, \mathrm{cm}$ and inlet flow rate $q= 22 \, \mathrm{cm^2/s}$, using an equal volume mixture of size-bidisperse $d_A/d_B=2$ particles with large particle diameter $d_A=2 \, \mathrm{mm}$. The slope of the fit lines, $S_{A,B} \approx -S_{B,A}$, is the segregation length scale, which depends on physical properties of particles but not mixture concentrations or flow kinematics. This technique has been used to calculate $S$ for mixtures of spherical particles bidisperse in either size~\cite{schlick2015} or density~\cite{xiao2016} and for cylindrical particle mixtures bidisperse in length~\cite{zhao2017}.}
\label{percolationVelocityOldMethod}
\end{center}
\end{figure}

The percolation velocity, $w_{s,i}$, in Eq.~(\ref{continuumModel}) is the local mean normal velocity of species $i$ relative to the local mean velocity, and reflects the segregation of the two species, denoted as $A$ and $B.$ It depends on the local concentration of species, $c_i,$ and the local shear rate, $\dot{\gamma}$,~\cite{savageLun1988,fan2014} in addition to particle mixture properties such as the particle size ratio, $R_{S}=d_{A}/d_{B},$ for size-bidisperse particles~\cite{schlick2015}, or the particle density ratio, $R_{D}=\rho_{A}/\rho_{B},$ for density-bidisperse particles~\cite{xiao2016}, where $d_i$ and $\rho_i$ are the species diameter and density, respectively. An example of the dependence of the percolation velocity on species concentration and shear rate for a DEM simulation of a bidisperse mixture is shown in Fig.~\ref{percolationVelocityOldMethod}. To obtain this data, the thin flowing layer at the top of the bounded heap flow (see Fig.~\ref{quasi2DBoundedHeapGeometryFigure}) is subdivided into a grid of small rectangular bins in which $c_i$, $\dot{\gamma}$, and $w_{s,i}$ are averaged over a 2\,s interval. All three values vary substantially with location (streamwise and depthwise) in the flowing layer, but  when $w_{s,i}$ is plotted vs.\ $\dot{\gamma}(1-c_i),$ the percolation velocity data is reasonably well characterized as a linear function:
\begin{equation}\label{percolationVelocity}
w_{s,i}=S \dot{\gamma}(1-c_i),
\end{equation}
where the segregation coefficient, $S$, relates the percolation velocity of individual species to the local flow conditions. We note that although a percolation velocity model quadratically dependent on local concentration~\cite{vanderVaart2015,jones2018} is more accurate in certain situations (e.g., small concentrations of one species), the linear model is sufficiently accurate for this study. 

This method of calculating $S$ from direct measurements of percolation velocity in the flowing layer has been used to develop correlations for $S$ as a function of the size ratio for mixtures of size-bidisperse spheres~\cite{schlick2015,jones2018} and rods~\cite{zhao2017}, and the density ratio of density-bidisperse spheres~\cite{xiao2016} for mm-sized particles in DEM simulations. It is simple to measure $S$ in DEM simulations where the conditions at every point in the flow are known. It is quite challenging to measure $S$ experimentally using this direct approach, which motivates the alternative approach to measure $S$ proposed in this paper.

In addition to $w_{s,i},$ the other key parameter in Eq.~(\ref{continuumModel}) is the diffusion coefficient, which depends on the local particle size and shear rate as $D = C_{D} \dot{\gamma} \bar{d}^{2}$~\cite{bridgwater1980,utter2004}, where $\bar{d}$ is the volume-based mean particle diameter [for equal volume mixtures of size-bidisperse particles $\bar{d}=(d_L+d_S)/2$] and $C_{D}$ depends on particle material properties and geometry and typically has a value (determined from local measurements of particle mean squared displacement in dense granular flow) around 0.1~\cite{utter2004,fan2015}. Section~\ref{modelSensitivity} shows that diffusion has only a minimal effect on the model solution in the bounded heap geometry, so we use $C_{D}=0.1$, consistent with previous research~\cite{fan2015,xiao2019}. 

\section{Segregation coefficient from concentration profiles}\label{backfitMethod}

The solution of the continuum model for segregation [Eq.\ (\ref{continuumModel})] provides the concentration of both species throughout the entire flowing layer. However, only the concentration at the bottom of the flowing layer is used in our approach because it corresponds to the particles deposited on the heap, which is easily measured in experiment. In the quasi-2D bounded heap flow geometry, solving the continuum model requires the segregation coefficient, $S$, diffusion coefficient, $D$, and flowing layer thickness, $\delta,$ as discussed above. The approach described here estimates $S$ from the concentration profile in the static portion of the heap and obtains $\delta$ and $D$ through other means.

\subsection{Sensitivity to $S$, $D$, and $\delta$}\label{modelSensitivity}

To justify our approach, we begin by determining which continuum model parameters ($S$, $D$, and $\delta$) are most crucial to the parameter estimation method. As a first step, consider the dimensionless quantities $\Lambda=|S|L/\delta^{2}$ and $Pe=2q\delta/(DL)$ characterizing granular segregation~\cite{fan2014,schlick2015}. These parameters are, respectively, the ratio of an advection timescale to a segregation timescale and the ratio of a diffusion timescale to an advection timescale. The product of the two quantities $\Lambda Pe = 2qS/(\delta D)$ is then the ratio of a diffusion timescale to a segregation timescale. For typical flows of size-bidisperse granular mixtures in a quasi-2D heap, these values are on the order of $\Lambda \sim 10^{-1}$ and $Pe \sim 10^2$, so we expect the dominant competition to be between segregation and advection, with the importance of diffusion being smaller.

To test the sensitivity of the segregation state to $S$, $D$, and $\delta$ in an example quasi-2D bounded heap geometry, we use the continuum model to quantify the deviation of the deposited large particle streamwise concentration profile, $c_L(x)$, from a reference concentration profile, $c_{L,ref}(x)$, using the root-mean-squared-deviation: $RMSD= \sqrt{1/n \sum_{i=1}^{n} [c_{L,ref}(x_i) - c_{L}(x_i)]^{2}}$, where $n$ is the number of sampling bins. In other words, we find $RMSD$ as a function of $S$, $C_D$, and $\delta_0$ (the flowing layer thickness measured upstream) for reference values, $S_{ref}$, $C_{D,ref}$, and $\delta_{0,ref}.$ $S$ and $C_D$ control the rates of segregation and diffusion in the problem, respectively. $\delta$ enters the continuum model as a critical parameter in the velocity profile, which controls the advection, as well as the segregation and diffusion, via the shear rate, $\dot{\gamma} \sim u/\delta,$ and the characteristic distance that segregating particles travel before reaching the top or bottom of the flowing layer. 

\begin{figure}[!ht]
\begin{center}
\includegraphics[width=\columnwidth]{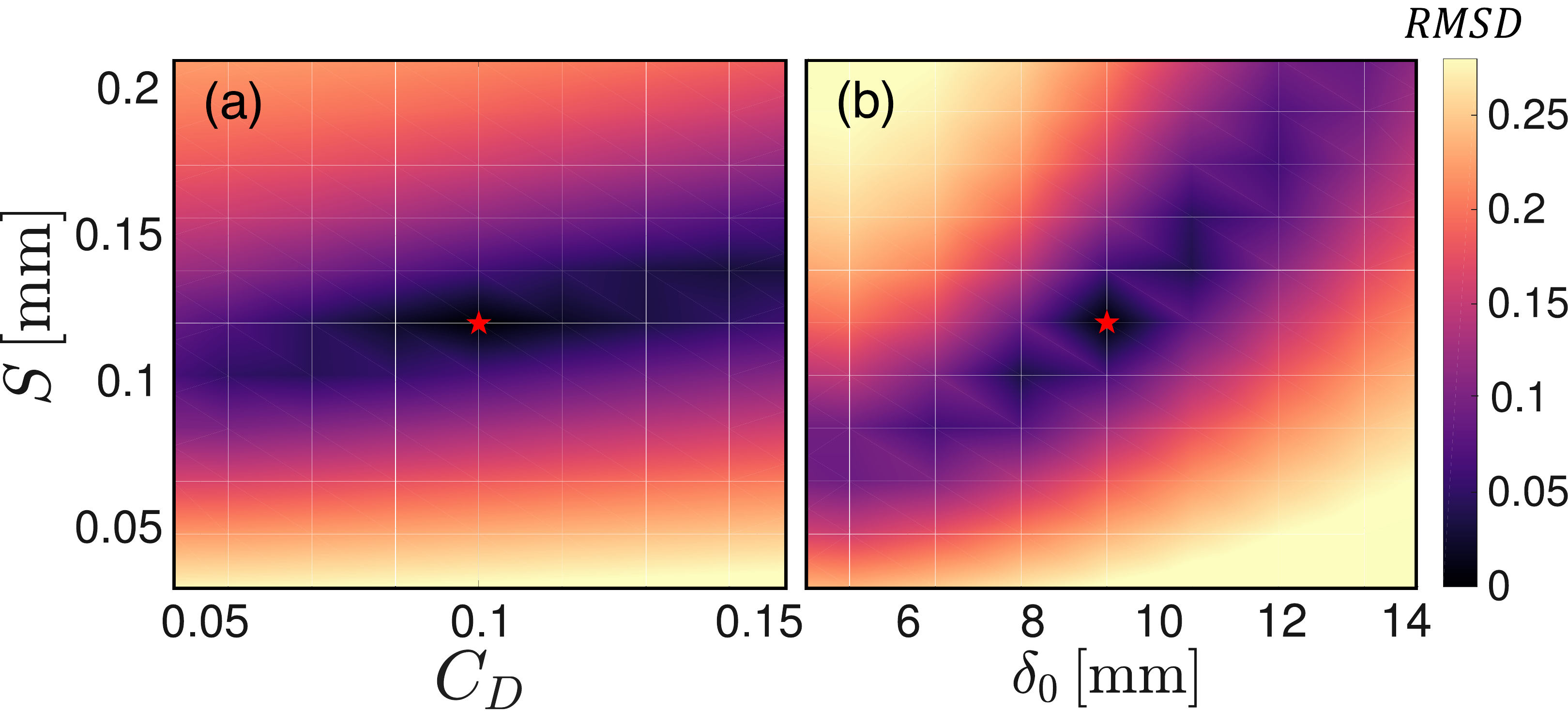}
\caption{Root mean squared deviation, $RMSD$, for model solutions of Eq.~(\ref{continuumModel}) at combinations of (a) segregation coefficient, $S$, and diffusion leading coefficient, $C_D$, and (b) $S$ and upstream flowing layer thickness, $\delta_0$. The reference model parameters ($S_{ref}= 0.12 \, \mathrm{mm}$, $C_{D,ref}= 0.1 \, \mathrm{mm^2/s}$, and $\delta_{0,ref}=9.2 \, \mathrm{mm}$) are denoted on each plot by a red star ($\star$) and correspond to a quasi-2D bounded heap flow with flowing layer length $L= 50 \, \mathrm{cm}$ and an equal volume mixture of size-bidisperse $d_{L}/d_{S}=2$ particles with mean diameter $\bar{d}=1 \, \mathrm{mm}$ and 2D inlet flow rate $q=10 \, \mathrm{cm^2/s}$~\cite{schlick2015}. An exponential velocity profile [see Eq.~(\ref{flowingLayerKinematicsStreamwise})] is assumed in all cases, $\delta$ is assumed to vary along the streamwise direction due to decreasing local flow rate, i.e.\ $\delta \sim q^{0.15}$, and the diffusion coefficient is calculated locally as $D= C_D \dot{\gamma} \bar{d}^2$, where $\dot{\gamma}$ is the local shear rate and $\bar{d}$ is the mean particle diameter.}
\label{parameterSensitivity}
\end{center}
\end{figure}

Figure~\ref{parameterSensitivity}(a) shows the $RMSD$ of $c_L$ as $S$ and $C_D$ are varied from the reference values $S_{ref}=0.12 \, \mathrm{mm}$ and $C_{D,ref}= 0.1 \, \mathrm{mm^2/s}$ (central red star) for constant $\delta_0=\delta_{0,ref}$. The reference values are for a size ratio $d_L/d_S=2$ mixture of mean diameter $\bar{d}= 1 \, \mathrm{mm}$ spherical particles in a flowing layer of length $L=50 \, \mathrm{cm}$ and inlet feed rate rate $q_0=10 \, \mathrm{cm^2/s}$. Varying $C_D$ at constant $S$ results in relatively small deviation from the reference profile, while varying $S$ at constant $C_D$ results in relatively large deviation, implying that the flow in the quasi-2D bounded heap (i.e., dense, and with a thin flowing layer) lies in a regime where segregation dominates diffusion~\cite{fan2014} for even modest size ratios (e.g., $d_L/d_S=2$ tested here). Consequently, the accuracy of the diffusion parameter prediction is relatively unimportant in the parameter estimation method in the quasi-2D heap geometry. This conclusion is consistent with previous research in which sensitivity of model predicted segregation to $D$ is minimal in bounded heap~\cite{fan2014} and chute~\cite{tunuguntla2016} geometries. Consequently, we treat diffusion as a known quantity, $D=0.1 \dot{\gamma}\bar{d}^2$, instead of estimating it using the measured concentration profile and the continuum model [Eq.~(\ref{continuumModel})].

The other model input, flowing layer thickness, $\delta$, is not arbitrarily adjustable since it must accurately reproduce velocity profiles in simulation or experiment. $\delta$ sets the shear rate and determines how far the downward segregating species moves before reaching the bottom of the flowing layer and settling out onto the deposited heap. Figure~\ref{parameterSensitivity}(b) shows the $RMSD$ of $c_L$ vs.\ $S$ and $\delta_0$ for reference values $S_{ref}=0.12 \, \mathrm{mm}$ and $\delta_{0,ref}=9.2 \, \mathrm{mm}$ (red star) for constant $D=D_{ref}= 0.1 \dot{\gamma} \bar{d}^2$. Small changes in $\delta_0$ at constant $S,$ and vice versa, lead to relatively large deviations in the model predictions. Since the solution of the continuum model is sensitive to changes in both $S$ and $\delta$, and since similar solutions [the diagonal ``valley" of low error in Fig.~\ref{parameterSensitivity}(b)] exist at various combinations of $S$ and $\delta$, it is necessary to provide one of them as an input, rather than estimating them both simultaneously from the deposited concentration profiles in a bounded heap experiment. Because $\delta$ is an extrinsic parameter that is relatively easy to obtain experimentally, we use the measured value in the parameter estimation method to determine $S$. Furthermore, since $S$ along the ``valley" of low deviation in Fig.~\ref{parameterSensitivity}(b) increases super-linearly with $\delta$, precisely determining flowing layer thickness from measured kinematics is a key component of the segregation coefficient estimation process. Generating accurate estimates of $\delta$ in experiment is explored in detail later in Section~\ref{wallSeparationDelta} and also in Part~II of this work~\cite{fry2019b}.

\subsection{Parameter estimation method}\label{backfitMethodOverview}

\begin{figure*}[ht]
\begin{center}
\includegraphics[width=\textwidth]{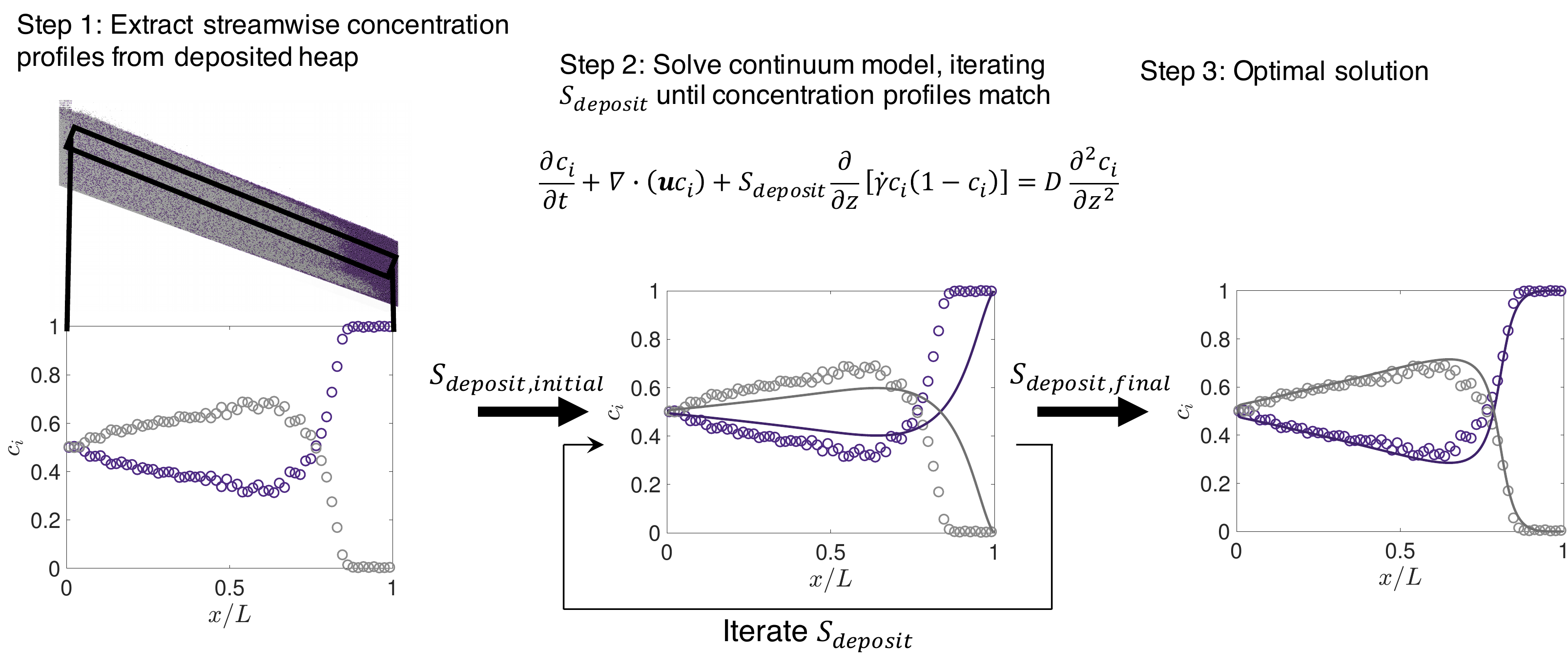}
\caption{Overview of $S$ estimation method. Step 1: Quasi-2D bounded heap (here, $W= 60\,\mathrm{cm}$, $T/d_L= 5.5$) is formed at constant feed rate, and the concentration of each species deposited in the heap is measured vs.\ streamwise location, $x/L$ (here we use an equal-volume mixture of $d_L/d_S=2$ particles at feed rate $q_0= 22 \, \mathrm{cm^2/s}$). Step 2: Continuum model is solved iteratively using an initial guess for the segregation coefficient (here, $S_{deposit,initial}=0.1 \, \mathrm{mm}$). Step 3: Best fit value (here, $S_{deposit,final}=0.21 \, \mathrm{mm}$) minimizes deviation between $c_i$ from the continuum model solution at the bottom of the flowing layer domain (curves) and $c_i$ in the deposited heap (circles). The velocity profile used in the continuum model is exponential, with locally varying flowing layer thickness, $\delta= \delta_0(1-x/L)^{0.15}$, where $\delta_0=2.4$\,mm, and locally varying diffusion coefficient, $D= 0.1 \dot{\gamma} \bar{d}^2$, where $\dot{\gamma}=\partial u/\partial z$ is the local shear rate calculated from the velocity profile, and the mean particle diameter is $\bar{d}= 1.65 \, \mathrm{mm}$. For comparison, $S$ measured in the flowing layer at this size ratio is $S_{flow}=0.20 \, \mathrm{mm}$~\cite{fan2014,schlick2015,xiao2016,zhao2017}.}
\label{backfitMethodPlot}
\end{center}
\end{figure*}

Having shown that the parameter estimation method is best implemented with $D$ taken from known correlations~\cite{utter2004,fan2015} and $\delta$ measured from experiments, while estimating $S$ from deposited concentration profiles, we now give a detailed description of the parameter estimation method, which is the key result of this paper, and test its efficacy using quasi-2D bounded heap simulation data.

A schematic of the $S$ estimation approach is shown in Fig.~\ref{backfitMethodPlot}. A particle mixture is fed into the left end of a quasi-2D heap (upper panel of Step 1) and concentrations of both species are extracted along the length of the deposited heap (lower panel of Step 1).  Bin averaged data is collected for the boxed portion of the heap located below the flowing layer and above the base of the heap where segregation is developing at the beginning of heap formation. 

In Step 2, the continuum model [Eq.~(\ref{continuumModel})] is solved numerically with an initial guess for the segregation coefficient, $S_{deposit,initial}=0.1 \, \mathrm{mm},$ and the known flow parameters, namely the exponential streamwise and normal velocity profiles [Eqns. (\ref{flowingLayerKinematicsStreamwise}, \ref{flowingLayerKinematicsNormal})], the flowing layer thickness measured in the flowing portion of the simulation during heap formation [$\delta(x)= \delta_0 (1-x/L)^{0.15}$, with $\delta_0=2.4 \, \mathrm{mm}$], and the diffusion coefficient ($D= 0.1 \dot{\gamma} \bar{d}^2$). The continuum model solution for the concentration at the bottom of the flowing layer (i.e.\ $z=-\delta$), which represents the particles deposited on the heap, is compared to the DEM heap data, as shown in Step 2. Then, $S$ is iteratively adjusted to minimize the error between the DEM simulation concentration profile and the profile predicted by the continuum model. The continuum model is solved numerically in MATLAB using the built-in differential equation solver $pdepe$ with grid resolution $n_x=200$ by $n_z=200$ and an optimization tolerance $optimset=10^{-6}$. The optimization method to find the best fit solution to the continuum model by modifying $S$ is performed in MATLAB using the built-in function $lsqnonlin$, which is an implementation of a non-linear Trust Region Reflective Least Squares algorithm~\cite{coleman1996,coleman1994} that does not calculate the analytical Jacobian of the objective function. The optimization problems solved in this study generally completed in about 50 function evaluations.

The final result (Step 3) is the $S_{deposit}$ value that minimizes the difference between the deposited concentration profiles calculated using the continuum model and measured from the DEM simulation. In Part~II of this work, concentration profiles from DEM simulations are replaced by concentration profiles from heap flow experiments. 

\subsection{Method validation}\label{methodValidation}

To validate the method, we compare the segregation coefficient, $S_{deposit}$, estimated from concentration profiles in the deposited portion of a quasi-2D bounded heap in DEM simulations (as shown in Fig.~\ref{backfitMethodPlot}) to the segregation coefficient, $S_{flow}$, measured in the flowing layer of the same DEM simulations (as shown in Fig.~\ref{percolationVelocityOldMethod}). A streamwise-varying flowing layer thickness, $\delta(x)=\delta_0 (1-x/L)^{0.15}$, where the maximum flowing layer thickness, $\delta_{0}$, is measured in the upstream portion of the DEM simulations (see Section~\ref{quasi2DBoundedHeap}), and a locally-varying diffusion coefficient, $D= 0.1 \dot{\gamma} \bar{d}^2$, are used to solve the continuum model. For all cases tested here, the sidewall gap is $T/d_L=5.5$. To solve the continuum model, particles are assumed to be uniformly mixed in the feed zone. 

\begin{table}[ht]
\caption{Simulation conditions for Fig.~\ref{fig:methodValidation} (equal-volume mixtures).}\label{simulationConditions}
\def\arraystretch{1.25}
\begin{tabular}{lcccccc}
\hline
\hline
   $d_L/d_S$& $d_L$ & $d_S$  & $\bar{d} \, \mathrm{[mm]}$ & $q \, \mathrm{[cm^2/s]}$ & $W \, \mathrm{[m]}$ &\\[1pt]
    \hline
       1.25 & 2.0 & 1.6 & 1.8 & 7.5 & 0.31&\\
       1.25 & 2.0 & 1.6 & 1.8 & 12.6 & 0.31&\\
       1.25 & 2.2 & 1.76 & 1.98 & 46.8 & 0.6 &\\
       1.5 & 2.0 & 1.33 & 1.67 & 12.0 & 0.31&\\
       1.5 & 2.2 & 1.47 & 1.83 & 23.4 & 0.6 &\\
       1.5 & 2.2 & 1.47 & 1.83 & 45.0 & 0.6 &\\
       1.5 & 4.0 & 2.67 & 3.33 & 83.0 & 1 &\\
       1.75 & 2.0 & 1.14 & 1.57 & 11.1 & 0.31&\\
       1.75 & 2.0 & 1.14 & 1.57 & 22.5 & 0.31&\\
       1.75 & 2.2 & 1.26 & 1.72 & 44.4 & 0.6 &\\
       2 & 2.0 & 1.0 & 1.5 & 8.4 & 0.31 &\\
       2 & 2.0 & 1.0 & 1.5 & 13.2 & 0.31&\\
       2 & 2.0 & 1.0 & 1.5 & 17.1 & 0.31&\\
       2 & 2.0 & 1.0 & 1.5 & 21.3 & 0.31 &\\
       2 & 2.2 & 1.1 & 1.65 & 22.2 & 0.6 &\\
       2 & 2.2 & 1.1 & 1.65 & 43.2 & 0.6 &\\ 
       2 & 4.0 & 2.0 & 3.0 & 79.0 & 1 &\\
       2.5 & 2.0 & 0.8 & 1.4 & 10.2 & 0.31&\\
       2.5 & 2.0 & 0.8 & 1.4 & 14.4 & 0.31&\\
       3 & 2.0 & 0.67 & 1.33 & 10.2 & 0.31&\\
         \hline
         \hline
  \end{tabular}
\end{table}

\begin{figure}[ht]
\begin{center}
\includegraphics[width=\columnwidth]{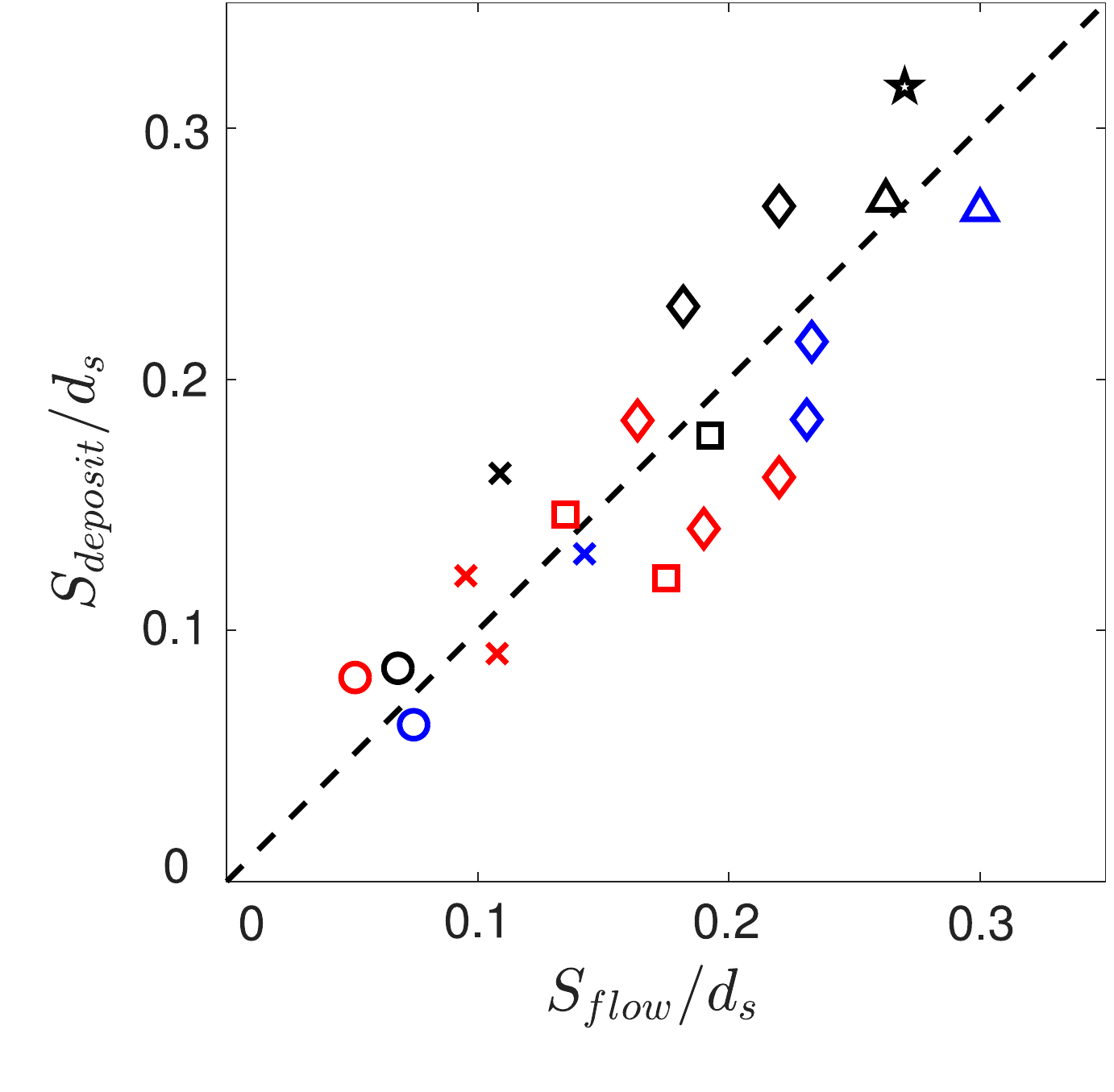}
\caption{Comparison of segregation coefficient estimated from deposited species concentrations, $S_{deposit}$, with segregation coefficient measured from particle velocity within the flowing layer, $S_{flow}$. Symbols from simulations detailed in Table~\ref{simulationConditions} with varying size ratio ($\circ$ -- $d_L/d_S=1.25$; $\times$ -- $d_L/d_S=1.5$; $\Box$ -- $d_L/d_S=1.75$; $\Diamond$ -- $d_L/d_S=2$; $\bigtriangleup$ -- $d_L/d_S=2.5$; $\bigstar$ -- $d_L/d_S=3$), and colors from simulations with varying rise velocity, $v_r=q_0/W$ (black -- $v_r< 4 \, \mathrm{mm/s}$; blue -- $4 \, \mathrm{mm/s} \leq v_r \leq 7 \, \mathrm{mm/s}$; red -- $v_r > 7 \, \mathrm{mm/s}$). Gap width $T/d_L=5.5$ for all simulations, and the inlet mixture ratio is 50\%-50\% (by volume). Mean particle diameter varies over the range $1.33 \, \mathrm{mm} \leq \bar{d} \leq 3.33 \, \mathrm{mm}$. Flowing layer thickness  $\delta=\delta_0 (1-x/L)^{0.15}$ and diffusion coefficient, $D= 0.1 \dot{\gamma} \bar{d}^2$, are inputs to the model in each case, with maximum flowing layer thickness $\delta_0$ taken from the upstream portion of the heap. }
\label{fig:methodValidation}
\end{center}
\end{figure}

To rule out systematic bias in the method and quantify its accuracy, we compare results at various system and mixture conditions (see Table~\ref{simulationConditions}). In each simulation, particle size ratio, $d_L/d_S$, and large particle size, $d_L$, are set and then used to calculate the small particle diameter, $d_S$, and volume based mean particle diameter, $\bar{d}$. The segregation coefficient estimated from the deposited streamwise concentration profiles, $S_{deposit}$, is plotted in Fig.~\ref{fig:methodValidation} vs.\ $S_{flow}$ measured from particle velocity data within the flowing layer. The two approaches are in good agreement over the broad range of conditions tested. Deviations are distributed evenly above and below a unit slope line (i.e.\ perfect correlation). 

In the figure, symbols and colors reflect different size ratios and heap rise velocities, $v_r=q_0/W$. It is evident that $S$ increases with size ratio, as expected. More interesting is  that $S_{deposit}$ is slightly overestimated for low $v_r$ (black) and slightly underestimated for moderate $v_r$  (blue) and some high $v_r$ (red). This is likely a result of two secondary segregation mechanisms that are not included in the continuum model: initial segregation in the feed zone and segregation due to small particles bouncing down the surface of the heap~\cite{fan2012}. Feed zone segregation (which increases $S_{deposit}$ relative to $S_{flow}$), measured as the mean deviation from $c_L=0.5$ in the normal direction at the feed zone exit, increases with decreasing inlet flow rate (i.e., lower $v_r$) and increasing size ratio. Bouncing of small particles at the heap surface (which decreases $S_{deposit}$ relative to $S_{flow}$) increases with increasing $q_0$ (i.e., higher rise velocity) and increasing size ratio (as previously reported~\cite{fan2012}). The sum of these secondary mechanisms, though, has a relatively small effect on the prediction of $S$ for the cases tested. Moreover, the key result of this paper is that the parameter estimation approach outlined in Fig.~\ref{backfitMethodPlot} can determine $S$ from deposited species concentration profile data, whether that data comes from DEM simulations, as in this paper, or from actual experiments, as in Part~II of this study\cite{fry2019b}.

\section{Considerations for practical implementation} \label{flowConditions}

Since this method of determining $S$ from deposited concentration profiles is ultimately intended for practical application, it must be applicable in cases where flow kinematics can only be measured at the sidewall, the free surface, or both. This makes spanwise variation in the velocity profile problematic, because it is not possible to determine the velocity profile in the bulk away from the wall during flow, except at the free surface. Furthermore, spanwise variations in the velocity can lead to spanwise variations in the concentration of particles deposited in the heap. It is quite challenging to sample the deposited heap effectively in the spanwise direction, and streamwise samples would smear variations in local concentration profiles in the spanwise direction. Since streamwise velocity profile variation in the spanwise direction is due to frictional sidewalls~\cite{jop2005,taberlet2003,brodu2013,baker2016}, we study next the effect of varying the gap between the sidewalls on local segregation in bounded heap flow.

\subsection{Influence of spanwise gap on flowing layer depth}\label{wallSeparationDelta}

\begin{figure}[!ht]
\begin{center}
\includegraphics[width=3in]{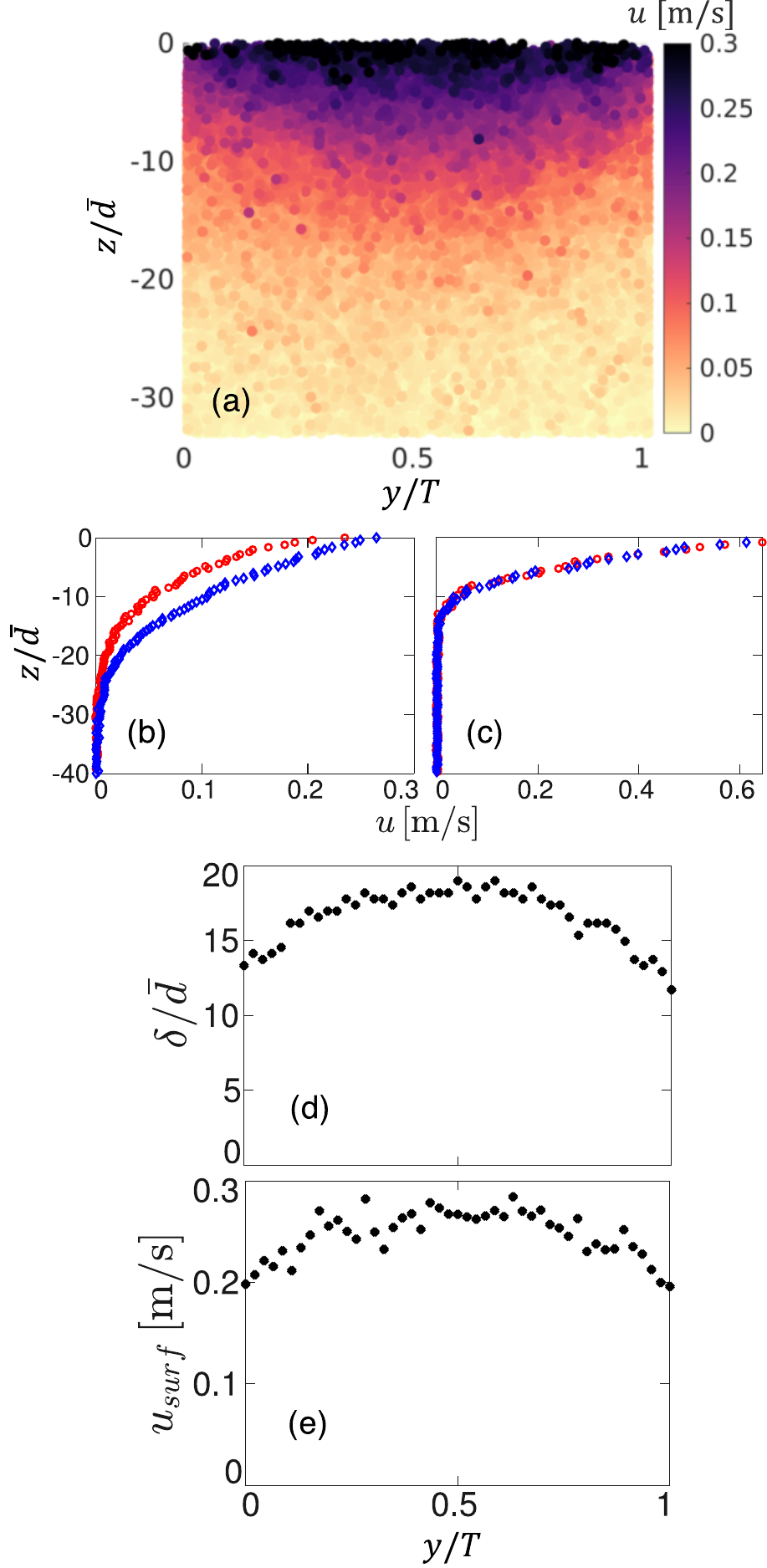}
\caption{(a) Instantaneous streamwise velocity of individual particles (filled circles), $u$, vs.\ normalized spanwise, $y/T$, and vertical, $z/\bar{d}$, location for all particles at streamwise location $x=L/4$ in simulation with gap width $T/d_L=39.6$, showing increased flowing layer thickness away from the walls. Streamwise velocity averaged across  10\% of the gap closest to the walls (red circles) and across 10\% of the gap at the center (blue diamonds) vs.\ depth normalized by mean particle diameter, $z/\bar{d}$ for (b) $T/d_L=39.6$ and (c) $T/d_L=5.5$. (d) Flowing layer thickness normalized by mean particle diameter, $\delta/\bar{d}$, and (e) streamwise surface velocity, $u_{surf}$, vs.\ spanwise position normalized by gap width, $y/T$ for all particles at $x=L/4$ in a simulation with $T/d_L=39.6$. All data taken from cases with $W=1 \, \mathrm{m}$, $d_L/d_S=1.5$, $d_L=4 \, \mathrm{mm}$, and $v_r= q_0/W= 8 \, \mathrm{mm/s}$.}
\label{heapSpanwiseVariation}
\end{center}
\end{figure}

As shown previously in granular chute experiments~\cite{jop2005}, as the gap between sidewalls, $T$, increases, flow properties including the surface velocity, $u_{surf}$, and the thickness of the flowing layer, $\delta$~\cite{jop2005}, vary substantially across the width of the gap. For example, in Fig.~\ref{heapSpanwiseVariation}(a) the instantaneous streamwise velocity for all particles at streamwise location $x=L/4$ is displayed in a plane perpendicular to the flow direction. At any depth, velocity increases with increasing distance from the nearest wall. 

The velocity is plotted vs.\
depth from the free surface at the center-line and at the wall for a wide heap, $T/d_L=39.6$, in Fig.~\ref{heapSpanwiseVariation}(b) and for a narrow heap, $T/d_L=5.5$, in Fig.~\ref{heapSpanwiseVariation}(c). The flowing layer at both the sidewall and center-line  is noticeably deeper in the wider heap than the narrow heap. Furthermore, in the wide heap the centerline surface velocity is higher and flowing layer thickness is deeper compared to the surface velocity and the flowing layer thickness at the wall. In comparison, there is almost no difference between the centerline and near wall velocity profiles in the narrow gap case. Since the mean particle size, $\bar{d}$, particle size ratio, $d_L/d_S$, heap length, $W$, and rise velocity, $v_r=q_0/W$ (and hence, also the 2D feed rate, $q_0$) are held constant between the narrow gap and wide gap cases, the change in flow behavior is a result of changing gap width. 

To further illustrate the variation across the wide gap, the flowing layer thickness and surface velocity  are plotted vs.\ spanwise location, $y/T$ in Fig.~\ref{heapSpanwiseVariation}(d,e). The variation of both quantities across the gap is significant (e.g., $\delta_{center}/\delta_{wall}>1.5$). As shown in the next section, the degree of segregation in the deposited layer depends sensitively on $\delta$, so segregation predictions made using $\delta$ measured at the wall in wide gaps will be inaccurate. 

\begin{figure}[!ht]
\begin{center}
\includegraphics[width=\columnwidth]{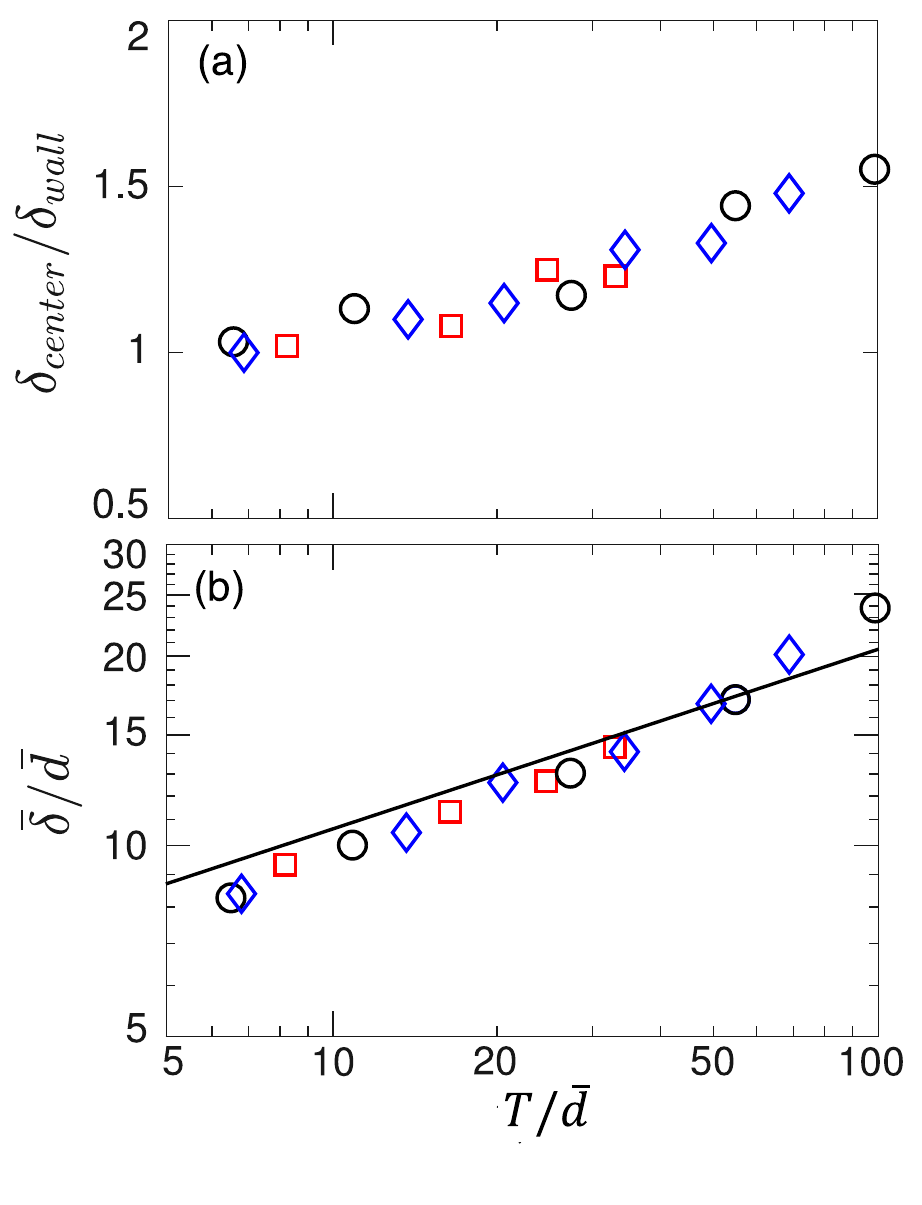}
\caption{(a) Ratio of mean flowing layer thickness in the middle 10\% of the gap to mean flowing layer thickness in the 10\% of the gap closest to the sidewalls, $\delta_{center}/\delta_{wall}$, and (b) span-averaged flowing layer thickness normalized by mean particle diameter, $\bar{\delta}/\bar{d}$, vs.\ spanwise gap width normalized by mean particle diameter, $T/\bar{d}$ for $d_L/d_S=1$ (black $\circ$), $d_L/d_S=1.5$ (blue $\Diamond$), and $d_L/d_S=2$ (red $\Box$). In (b) a slope $2/7$ line, as suggested for granular chute flow between bounding sidewalls~\cite{jop2005}, is provided for comparison. Data are taken at $x=L/2$ for DEM simulations of bounded heap flow with $W= 1 \, \mathrm{m}$, $2.7 \, \mathrm{mm} \leq \bar{d} \leq 4 \, \mathrm{mm}$, and $v_r= 1 \, \mathrm{mm/s}$.}
\label{deltaVsSeparation}
\end{center}
\end{figure}

Therefore, it is necessary to establish the gap thickness above which significant spanwise variation occurs in the streamwise velocity and flowing layer depth. Figure~\ref{deltaVsSeparation}(a) shows the ratio of the flowing layer thickness averaged over the center 10\% of the gap to the flowing layer thickness averaged over the 10\% of the gap closest to the sidewall, $\delta_{center}/\delta_{wall}$ vs.\ $T/\bar{d}$. All simulations are performed with the same system length and rise velocity, but at varying levels of size dispersity, $1<d_L/d_S<2$. A clear trend exists between the spanwise flowing layer depth variation and the spanwise gap width regardless of size ratio. For spanwise gap width less than about $10 \bar{d}$, variation in $\delta$ is minimal. At a gap width of about $25 \bar{d}$, $\delta_{center}$ is nearly 25\% thicker than $\delta_{wall}$, and at a  gap width of $100 \bar{d}$, $\delta_{center}$ is more than 50\% larger than $\delta_{wall}$. 

The influence of $\delta$ variation on segregation in quasi-2D bounded heap flow is captured by the dimensionless segregation-advection ratio, $\Lambda=|S|L/\delta^2,$ which indicates that increasing spanwise variation in $\delta$ leads to correspondingly larger variation in deposited concentration across the gap. In Fig.~\ref{deltaVsSeparation}(b), the span-averaged flowing layer thickness normalized by the mean particle diameter, $\bar{\delta}/\bar{d}$, is plotted vs.\ $T/\bar{d}$ for the same cases as in Fig.~\ref{deltaVsSeparation}(a). When $T/\bar{d}$ is small, the flowing layer is relatively thin ($\bar{\delta}/\bar{d}<10$). As $T/\bar{d}$ increases, the flowing layer thickness increases. Thus, the variation in $\delta$ across the wide gaps [Fig.~\ref{heapSpanwiseVariation}(a)] and the increase in $\delta$ with increasing gap width [Fig.~\ref{heapSpanwiseVariation}(b)] have the potential to reduce the effectiveness of the parameter estimation method described in Section~\ref{backfitMethodOverview} because of their effects on the segregation. 

As a side note, a reference line of slope 2/7, which corresponds to a power law scaling proposed in an experimental study of $\delta$ variation with gap width in granular chute flows~\cite{jop2005}, is overlaid on the data in Fig.~\ref{deltaVsSeparation}(b). Even though this relation was developed for gaps wider than 20 particle diameters in monodisperse flow, it extends to even narrower gaps ($\bar{\delta} / \bar{d}$ does not appear to level off as a function of $T/\bar{d}$, even for $T/\bar{d}<10$) as well as size disperse mixtures (i.e., different values of $d_L/d_S$). 

\subsection{Influence of spanwise velocity profile variation on deposited concentration profiles}

\begin{figure}[!ht]
\begin{center}
\includegraphics[width=\columnwidth]{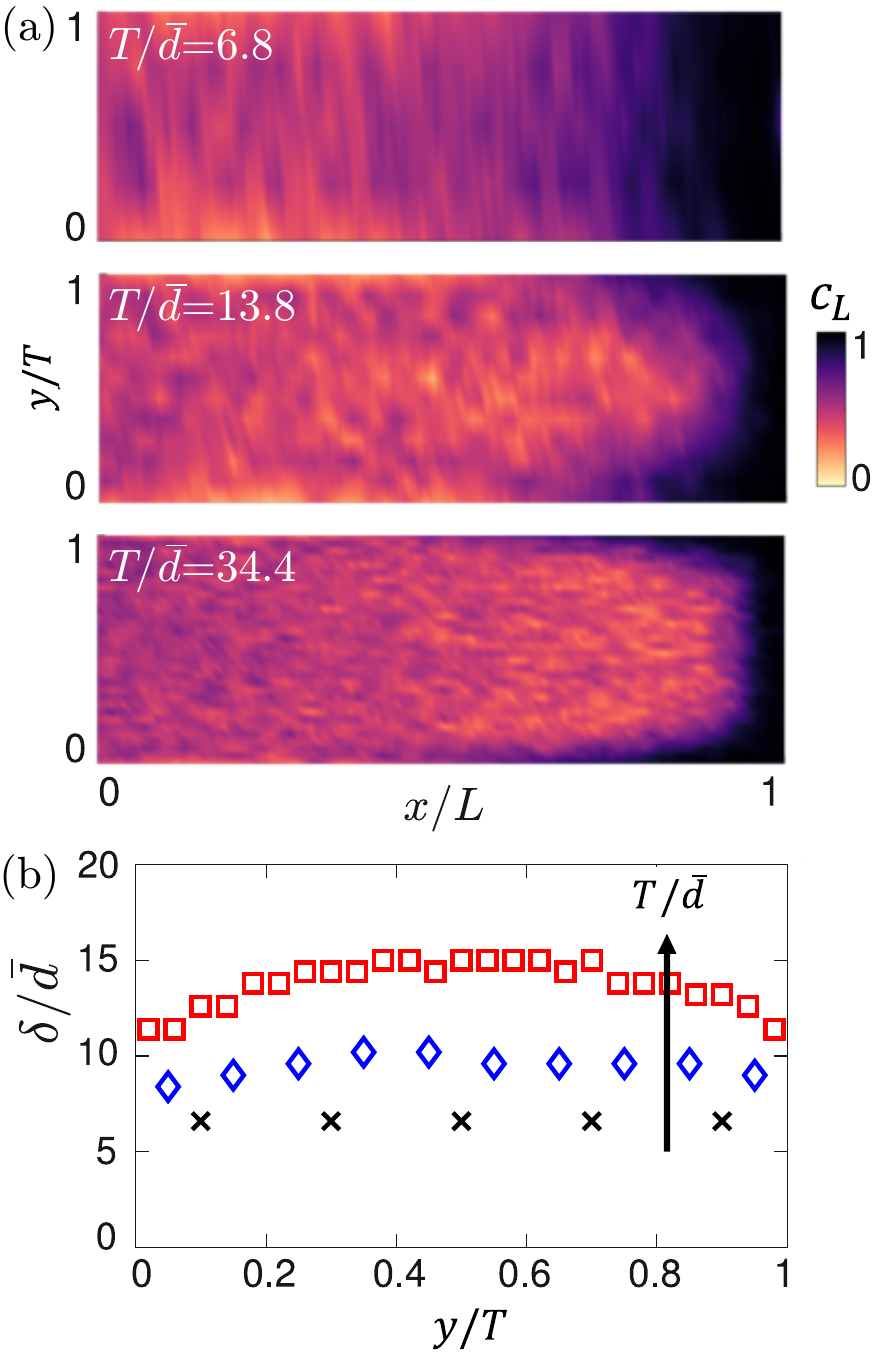}
\caption{(a) Top view of local large-particle-species concentration, $c_L$, at normalized spanwise, $y/T$, and streamwise, $x/L$, location in the deposited heap for simulations with varying spanwise gap width. Segregation increases toward the sidewalls with wider gaps, since flowing layer thickness decreases near the walls. (b) Flowing layer thickness normalized by mean particle diameter, $\delta/\bar{d}$, vs.\ $y/T$ for simulations in (a). Symbols represent varying gap width: $T/\bar{d}=6.8$ (black $\times$), $T/\bar{d}=13.8$ (blue $\Diamond$), and $T/\bar{d}=34.4$ (red $\Box$). Simulations performed with $W=1 \, \mathrm{m}$ and $v_r=q_0/W= 8 \, \mathrm{mm/s}$ for $d_L/d_S=1.5$ mixture with $\bar{d}=3.2 \, \mathrm{mm}$.}
\label{spanwiseVariationPlot}
\end{center}
\end{figure}

The previous section describes the variation in the velocity profile and flowing layer thickness across wide gaps and their potential to affect segregation in bounded heap flow, and thereby the implementation of the parameter estimation method described in Section~\ref{backfitMethodOverview}. In this section, we test the impact of spanwise variations in velocity on the deposited concentration profiles. Specifically, we perform DEM simulations of a size-bidisperse $d_L/d_S=1.5$ mixture in the bounded heap geometry with $W=1 \, \mathrm{mm}$ and $6.8<T/\bar{d}<34.4$. Figure~\ref{spanwiseVariationPlot}(a) shows a top-down view of the large particles concentration, $c_L,$ in the deposited heap vs. spanwise, $y/T$, and streamwise, $x/L$, location. The variation in segregation near the walls and at the center of the gap increases with increasing spanwise gap width, as is evident from the increased large particle concentration (dark) along the sidewalls, but not at the centerline.

As shown in Fig.~\ref{spanwiseVariationPlot}(b), the spanwise variation in $\delta$ increases with increasing gap width, consistent with Fig.~\ref{deltaVsSeparation}(a). The spanwise variation in the deposited concentration profiles at larger values of $T/\bar{d}$ in Fig.~\ref{spanwiseVariationPlot}(a) come about because a thicker flowing layer at the centerline means particles travel a greater distance before depositing on the heap, thereby reducing segregation locally. This is reflected in the local dimensionless advection-segregation parameter, $\Lambda=SL/\delta^2$, which decreases with increasing $\delta$ at the center of the gap.

\begin{figure}[!ht]
\begin{center}
\includegraphics[width=\columnwidth]{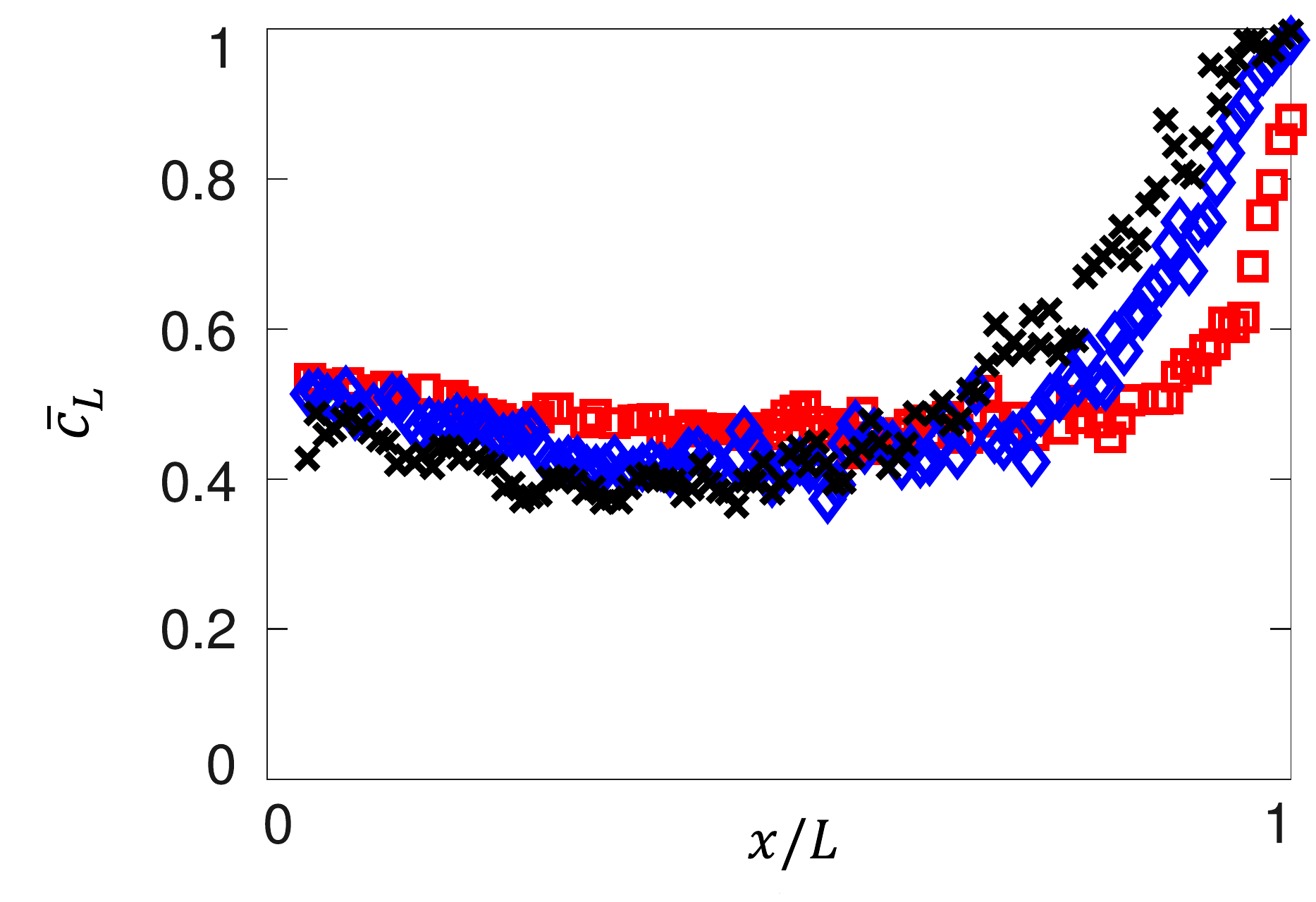}
\caption{Span-averaged large-particle concentration, $\bar{c}_L$, vs.\ streamwise location, $x/L,$ for simulations with varying gap width: $T/\bar{d}=6.8$ (black $\times$), $T/\bar{d}=13.8$ (blue $\Diamond$), and $T/\bar{d}=34.4$ (red $\Box$). ($d_L/d_S=1.5$, $\bar{d}=3.2 \, \mathrm{mm}$, $W=1 \, \mathrm{m}$, and $v_r=q_0/W= 8 \, \mathrm{mm/s}$.)}
\label{spanwiseSummationPlot}
\end{center}
\end{figure}

In addition to increasing the spanwise concentration variation, increasing gap width also leads to a deeper mean flowing layer overall, which is evident in Figs.~\ref{deltaVsSeparation}(b) and~\ref{spanwiseVariationPlot}(b). This reduces segregation throughout the flow, again due to reduced $\Lambda$ associated with larger $\delta$. This effect is evident in the concentration of large particles averaged across the span, $\bar{c}_L$, when plotted vs.\ streamwise position in the deposited heap in Fig.~\ref{spanwiseSummationPlot} for the different gap widths in Fig.~\ref{spanwiseVariationPlot}.  $\bar{c}_L$ is substantially lower  downstream (large $x/L$) for the largest gap case. Furthermore, in the upstream portion of the heap ($0.1 \leq x/L \leq 0.6$), $\bar{c}_L$ is closer to 0.5, indicating less segregation than with smaller spanwise gaps. Since increased segregation along the length of the heap improves the parameter estimation method, the enhanced segregation associated with narrower spanwise gap will improve the accuracy of the method, especially for particle mixtures with small size ratios, for which the segregation coefficient is small. 

In summary, there are two reasons to keep the gap between the sidewalls of the bounded heap small. First, a narrow gap decreases the overall flowing layer thickness [Figs.~\ref{deltaVsSeparation}(b) and~\ref{spanwiseVariationPlot}(b)], which results in stronger segregation that is easier to quantify (see Fig.~\ref{spanwiseSummationPlot}). Second, a narrow gap results in a spanwise invariant velocity field [Fig.~\ref{heapSpanwiseVariation}(c)] and flowing layer thickness [Figs.~\ref{deltaVsSeparation}(a) and~\ref{spanwiseVariationPlot}(b)], both of which simplify implementing the parameter estimation method. Consequently, the spanwise gap should be maintained below $T/\bar{d} \approx 15$ when applying the parameter estimation method.

\subsection{Streamwise variation in diffusion and flowing layer thickness in the model}\label{localVariation}

We now return to the issue of the dependence of the diffusion coefficient and the flowing layer thickness on streamwise position in the flowing layer, first mentioned in Section~\ref{quasi2DBoundedHeap}. A previous study~\cite{fan2014} showed that even though the diffusion coefficient depends on the local shear rate, using a local diffusion coefficient results in only a slightly better match between the continuum model predictions and simulation and experimental results. Likewise, previous studies have assumed a constant flowing layer thickness~\cite{fan2014,schlick2015,xiao2016} and achieved a good match between model predictions using Eq.~(\ref{continuumModel}) and DEM simulation and experimental results.

To test the impact of local variation in $D$ and $\delta$ on the accuracy of the parameter estimation technique, we compare estimates of $S$ where $\delta$ and $D$ are constant with estimates where they vary locally. Due to mass conservation, surface velocity and shear rate decrease linearly with streamwise location for a constant $\delta$~\cite{fan2014,schlick2015}, while they decrease sub-linearly when $\delta$ decreases downstream. For the simulations presented in this paper, the decrease in $\delta$ with local flow rate, $q,$ is well described by $\delta(q) \sim q^{\alpha}$, where $\alpha=0.15$ (see Section~\ref{quasi2DBoundedHeap}). Based on mass conservation, the surface velocity varies as $u_{surf} \sim q^{1-\alpha}$ and the shear rate varies as $\dot{\gamma} \sim q^{1-2\alpha}$. For constant $\delta$, $\alpha=0$, which suggests that differences in predicted $S$ for fixed vs.\ varying $\delta$ should be minimal given the small value of $\alpha=0.15$ used here.

\begin{figure}[!ht]
\begin{center}
\includegraphics[width=\columnwidth]{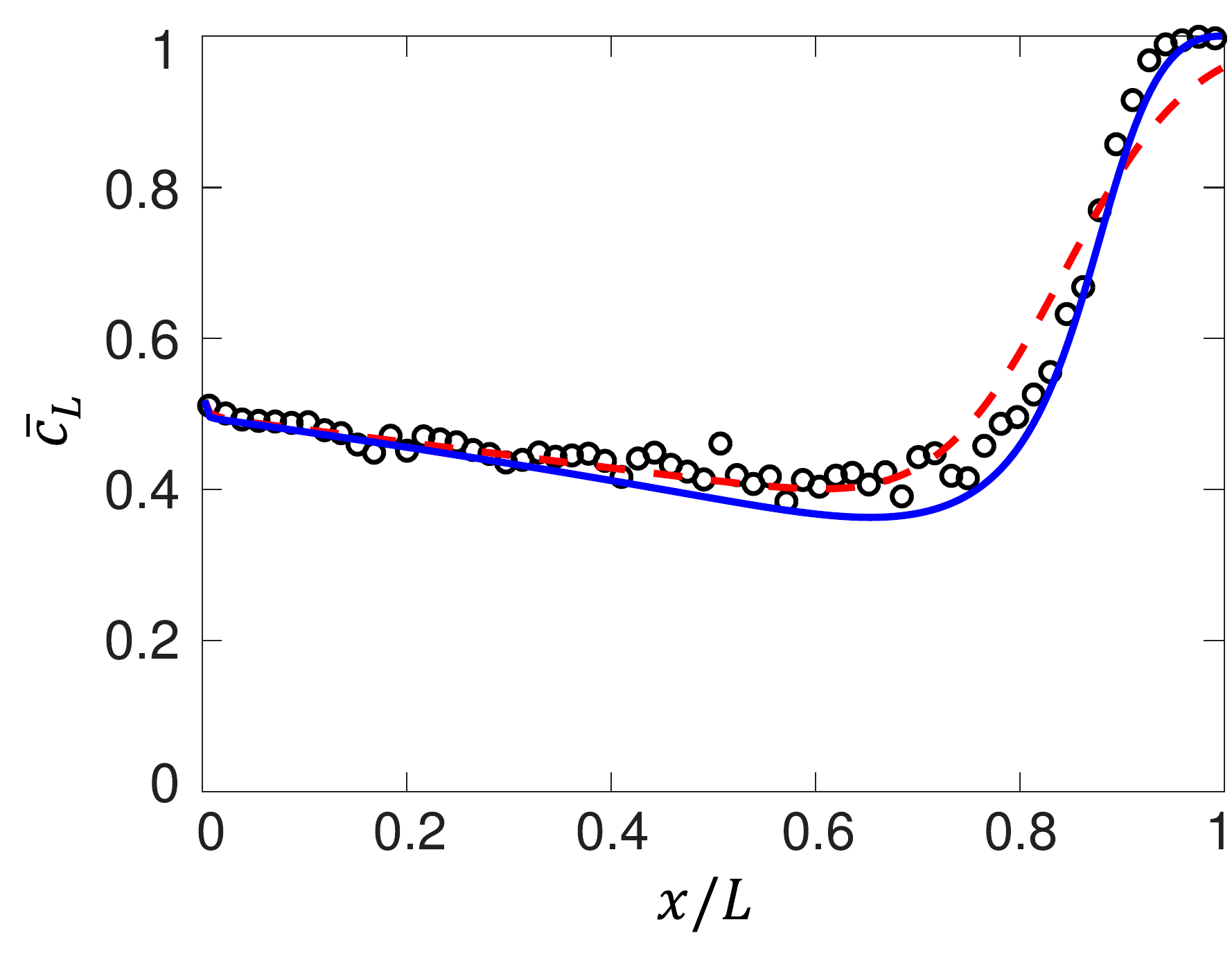}
\caption{Large species concentration, $c_{L}$, vs.\ streamwise position, $x/L$, for bounded heap simulation ($\circ$) compared to best fit solution of continuum model using constant diffusion coefficient, $D,$ and flowing layer thickness, $\delta=27$\,mm (red dashed curve) or locally varying $D(x,z)= \dot{\gamma}(x,z) \bar{d}^2,$ and $\delta(x)= 27.5 (1-x/L)^{0.15} \, \mathrm{mm}$ (blue solid curve) for $T/d_L=5.5$, $W= 60 \, \mathrm{cm}$, $\bar{d}= 1.72 \, \mathrm{mm}$, $d_L/d_S=1.75$, and $v_r= 7.4 \, \mathrm{mm/s}$.}
\label{localVsGlobalDelta}
\end{center}
\end{figure}

Best fits of the continuum model to the deposited large particle concentration profile using both constant and spatially varying $\delta$ and $D$ are plotted in Fig.~\ref{localVsGlobalDelta} along with the corresponding DEM simulation data. The fit using spatially varying $\delta$ and $D$ better matches the simulation results than the fit for constant $\delta$ and $D$, correctly predicting a higher level of segregation in the downstream region (where $\delta$ and $D$ are smaller) and following the DEM data slightly better for $x/L>0.7$. However, the difference between the two predictions is not large. 

\begin{figure}[ht]
\begin{center}
\includegraphics[width=\columnwidth]{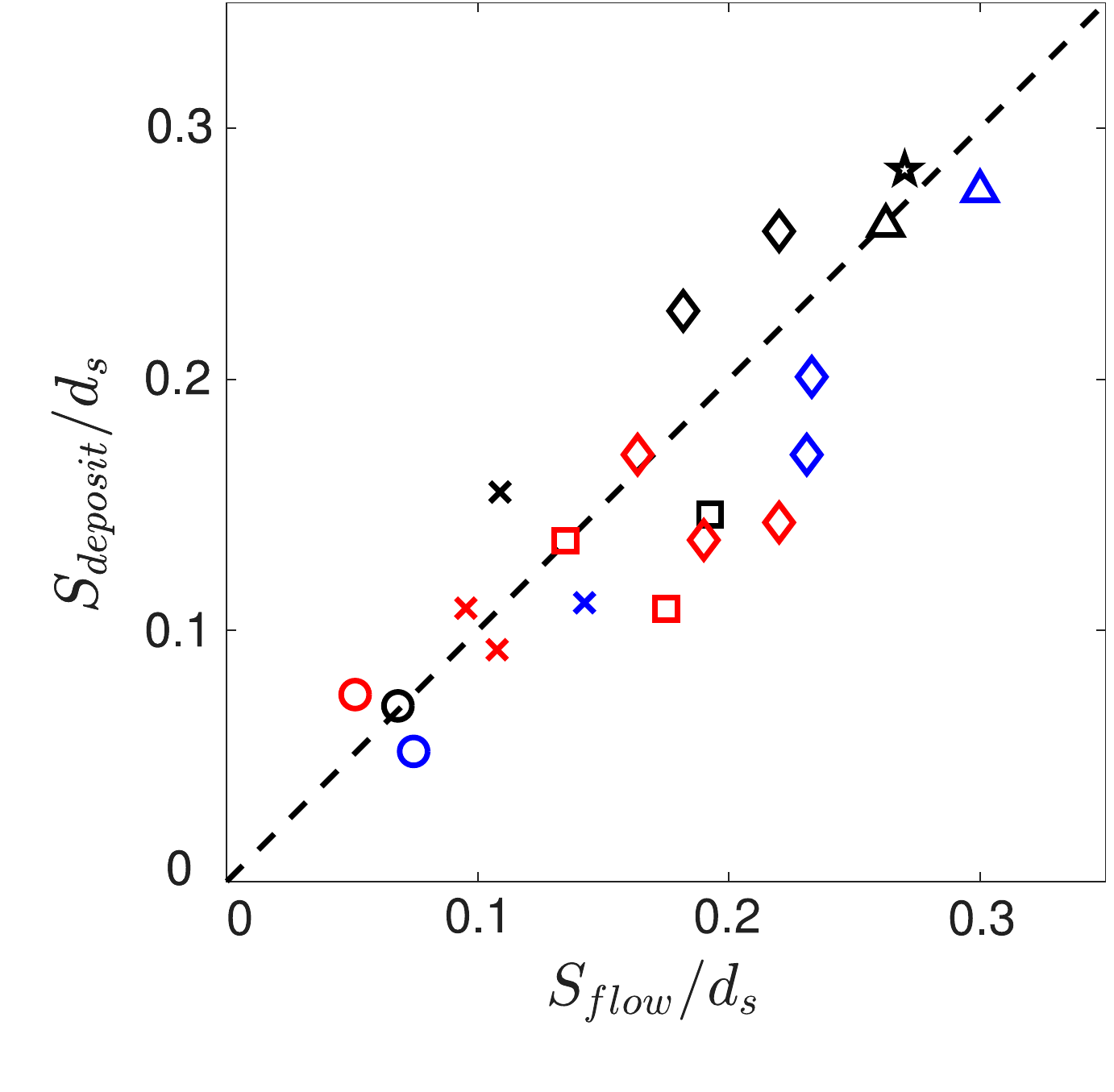}
\caption{Segregation coefficient estimated from deposited species concentrations, $S_{deposit}$, using continuum model with constant diffusion coefficient and flowing layer depth compared with segregation coefficient measured from particle velocity data within the flowing layer, $S_{flow}$. Symbols represent data from simulations detailed in Table~\ref{simulationConditions} with varying size ratio ($\circ$ -- $d_L/d_S=1.25$; $\times$ -- $d_L/d_S=1.5$; $\Box$ -- $d_L/d_S=1.75$; $\Diamond$ -- $d_L/d_S=2$; $\bigtriangleup$ -- $d_L/d_S=2.5$; $\bigstar$ -- $d_L/d_S=3$), and colors represent data from simulations with varying rise velocity (black -- $v_r< 4 \, \mathrm{mm/s}$; blue -- $4 \, \mathrm{mm/s} \leq v_r \leq 7 \, \mathrm{mm/s}$; red -- $v_r > 7 \, \mathrm{mm/s}$). Gap width $T/d_L=5.5$, inlet mixture ratio is 50\%-50\% (by volume) and mean particle diameter varies over the range $1.33 \, \mathrm{mm} \leq \bar{d} \leq 3.33 \, \mathrm{mm}$.}
\label{methodValidationGlobalKinematics}
\end{center}
\end{figure}

Perhaps more important is the value for the segregation coefficient, $S$, estimated using constant values for the flowing layer thickness and diffusion coefficient. A comparison analogous to that in Fig.~\ref{fig:methodValidation} for spatially varying parameters is shown in Fig.~\ref{methodValidationGlobalKinematics} for constant $D$ and $\delta$, where $\delta$ is measured in the upstream portion of each heap simulation and is $D$ taken from a known correlation~\cite{schlick2015}. The scatter is only slightly larger in Fig.~\ref{methodValidationGlobalKinematics} than Fig.~\ref{fig:methodValidation}, though the use of constant $D$ and $\delta$ appear to cause a slight under-prediction of $S_{deposit}$ for the conditions tested (i.e., more points in Fig.~\ref{methodValidationGlobalKinematics} fall below the line of unit slope than above it). Quantitatively, the mean absolute error for the segregation coefficient using the local flow rate dependent flow kinematics (data in Fig.~\ref{fig:methodValidation}) is approximately 5\% lower than the segregation coefficient fit using the streamwise-constant flowing layer thickness and diffusion coefficient for the range of flow conditions tested (data in Fig.~\ref{methodValidationGlobalKinematics}). 

Thus, using the local flow kinematics provides a modest improvement in continuum model prediction, both qualitatively, based on the shape of the fits, and quantitatively, based on the agreement between the deposit and flowing layer coefficients. Since implementing a local flow-rate dependent continuum model is straightforward, use of the local flow model is preferred. 

\section{Discussion}\label{discussion}
In order to apply a continuum model of mixing and segregation to industrially relevant granular mixtures, we have here, in Part I of this work, explored a method to estimate the segregation coefficient, $S$, using quasi-2D bounded heap simulations. The approach reverses the usual use of the advection-diffusion-segregation equation [Eq.~(\ref{continuumModel})].  Rather than using a known $S$ to predict segregated species concentrations, the parameter estimation method uses the measured deposited species concentration in a simple bounded heap flow to estimate $S$. Good agreement is achieved between the segregation coefficient estimated from the deposited heap and the segregation coefficient measured directly in the flowing layer of DEM simulations.

The practical challenge associated with the parameter estimation method we have described is to design an appropriate experiment that accounts for the issues that have been identified using these DEM simulations that validate the approach. The main concern is the gap between sidewalls. Increasing the gap results in an increase in the thickness and spanwise variation of the flowing layer. The scaling is independent of the size dispersity of the particles over the range of ratios examined, $1<d_L/d_S<2$. A wide gap results in two problems. First, the thicker flowing layer reduces the impact of segregation relative to advection in the flow because small particles have to percolate through a thicker flowing layer to deposit on the heap and the shear rate is lower. If the particles do not segregate by the end of the length of the heap, the degree of segregation is small and estimates of $S$ are inaccurate. A narrow gap is preferable, since the flowing layer is thinner and the shear rate is higher, which produces stronger segregation. The second problem is just as important. To attain accurate estimates of the model parameters to fit the concentration of particles deposited on the heap, the spanwise variation in velocity and species concentrations must be small. Hence, for both reasons, it is important that the gap between the sidewalls is less than about 15 mean particle diameters. 

We have shown here (Part I of our study) that species concentration and velocity data from a simple bounded heap together with a continuum segregation model have the potential to accurately determine the segregation coefficient of bidisperse particle mixtures. Experimental application of this approach will be of substantial value in cases where DEM simulations are impractical, such as for mixtures of non-spherical particles or where particle properties are challenging to quantify. In Part II of this study~\cite{fry2019b}, we perform heap segregation experiments and determine the segregation coefficient using the approach described here. The experimental challenges of accurately measuring the velocity profile based on wall and free surface observations and determining the  deposited species concentrations are described there in more detail.

\section*{Acknowledgements}
Funded by the Procter \& Gamble Company.

\appendix
\section{Discrete Element Method (DEM) simulations}\label{DEMSimulations}

In this study, quasi-2D bounded heap flows of granular materials are simulated using soft-sphere discrete element method (DEM) simulations~\cite{cundallStrack1979}. The benefit of simulations is that the motion of particles can be measured at \textit{all locations} during heap formation. Particles move according to Newton's laws of motion and collide with each other and with the system boundaries according to forces derived from a linear spring-dashpot models (normal forces) and a stick-slip model (tangential forces) that alternates between Coulomb sliding friction or a static contact spring-dashpot model depending on whether the contact reaches the sliding transition. 

Formally, the contact equations are $$\mathbf{f_{ij}^{n}}=\big[k_{n}\mathbf{\zeta}-2\gamma_{n} m_{eff}(\mathbf{V}_{ij} \cdotp \mathbf{\hat{r}}_{ij})\big]\mathbf{\hat{r}}_{ij}$$ for normal forces and $$\mathbf{f_{ij}^{t}}=\min{\big\{|k_{t} \beta-2\gamma_{t} m_{eff}(\mathbf{V}_{ij} \times \mathbf{\hat{r}}_{ij})|,|\mu \mathbf{F}_{ij}^{n}|\big\}}\mathrm{sgn}(\beta)\mathbf{\hat{s}}$$ for tangential forces. To model the tangential static friction force for non-slipping contact, the tangential displacement is given by $\beta(t)=\int_{t_{s}}^{t} V_{ij}^{t} dt$~\cite{schafer1996}, where $V_{ij}^{t}$ is the instantaneous tangential velocity between contacting particle surfaces, $t$ is the current time, and $t_{s}$ is the initial contact time. For sliding tangential contact, the friction coefficient is $\mu=0.4$. The normal collision parameters are calculated as $k_{n}=\big[(\pi/t_{c})^2+\gamma_n^2\big]m_{eff}$ and $\gamma_n=-\ln{(\varepsilon)}/t_{c}$, and the tangential parameters are calculated as $k_{t}=2/7k_{n}$ and $\gamma_{t}=2/7\gamma_n$, where $\varepsilon=0.8$ is the restitution coefficient, $m_{eff}=m_{1}m_{2}/(m_{1}+m_{2})$ is the effective mass, and $t_{c} = 2.5 \times 10^{-4} \, \mathrm{s}$ is the binary collision time. Walls in particle-wall collisions are modeled as flat frictional planes using the same contact equations as particle-particle collisions, with the wall sliding friction coefficient $\mu_{w}=0.4$, which was found in previous studies to produce a flowing angle of repose consistent with quasi-2D bounded heap experiments using glass particles~\cite{fan2013,xiao2016}. The integration scheme used is the symplectic Euler algorithm. For numerical stability, the integration timestep is $\Delta t=t_{c}/40$, as in previous publications~\cite{schlick2015}. 

\section*{References}
\bibliographystyle{elsarticle-num} 
\bibliography{fry_parameterEstimation}

\begin{thebibliography}{10}
\expandafter\ifx\csname url\endcsname\relax
  \def\url#1{\texttt{#1}}\fi
\expandafter\ifx\csname urlprefix\endcsname\relax\def\urlprefix{URL }\fi
\expandafter\ifx\csname href\endcsname\relax
  \def\href#1#2{#2} \def\path#1{#1}\fi

\bibitem{ottinoKhakhar2000}
J.~M. Ottino, D.~V. Khakhar, Mixing and segregation of granular materials,
  Annu. Rev. Fluid Mech. 32~(1) (2000) 55--91.

\bibitem{gray2018}
J.~M. N.~T. Gray, Particle segregation in dense granular flows, Annu. Rev.
  Fluid Mech. 50~(1) (2018) 407--433.
\newblock \href {http://dx.doi.org/10.1146/annurev-fluid-122316-045201}
  {\path{doi:10.1146/annurev-fluid-122316-045201}}.

\bibitem{meier2007}
S.~W. Meier, R.~M. Lueptow, J.~M. Ottino, A dynamical systems approach to
  mixing and segregation of granular materials in tumblers, Adv. Phys. 56~(5)
  (2007) 757--827.

\bibitem{umbanhowar2019}
P.~B. Umbanhowar, R.~M. Lueptow, J.~M. Ottino, Modeling segregation in granular
  flows, Annu. Rev. Chem. Biomol. Eng. 10 (2019) 129--153.

\bibitem{fan2014}
Y.~Fan, C.~P. Schlick, P.~B. Umbanhowar, J.~M. Ottino, R.~M. Lueptow, Modelling
  size segregation of granular materials: the roles of segregation, advection
  and diffusion, J. Fluid Mech. 741 (2014) 252--279.

\bibitem{hillTan2014}
K.~M. Hill, D.~S. Tan, Segregation in dense sheared flows: gravity, temperature
  gradients, and stress partitioning, J. Fluid Mech. 756 (2014) 54--88.

\bibitem{bertuola2016}
D.~Bertuola, S.~Volpato, P.~Canu, A.~C. Santomaso, Prediction of segregation in
  funnel and mass flow discharge, Chem. Eng. Sci. 150 (2016) 16--25.

\bibitem{bridgwater1985}
J.~Bridgwater, W.~S. Foo, D.~J. Stephens, Particle mixing and segregation in
  failure zones -- theory and experiment, Powder Tech. 41~(2) (1985) 147--158.

\bibitem{dolgunin1995}
V.~N. Dolgunin, A.~A. Ukolov, Segregation modeling of particle rapid gravity
  flow, Powder Tech. 83~(2) (1995) 95--103.

\bibitem{grayThornton2005}
J.~M. N.~T. Gray, A.~R. Thornton, A theory for particle size segregation in
  shallow granular free-surface flows, Proc. R. Soc. A 461~(2057) (2005)
  1447--1473.

\bibitem{fan2013}
Y.~Fan, P.~B. Umbanhowar, J.~M. Ottino, R.~M. Lueptow, Kinematics of
  monodisperse and bidisperse granular flows in quasi-two-dimensional bounded
  heaps, Proc. R. Soc. A 469~(2157) (2013) 20130235.

\bibitem{schlick2015}
C.~P. Schlick, Y.~Fan, A.~B. Isner, P.~B. Umbanhowar, J.~M. Ottino, R.~M.
  Lueptow, Modeling segregation of bidisperse granular materials using physical
  control parameters in the quasi-2d bounded heap, AIChE J. 61~(5) (2015)
  1524--1534.

\bibitem{schlick2015b}
C.~P. Schlick, Y.~Fan, P.~B. Umbanhowar, J.~M. Ottino, R.~M. Lueptow, Granular
  segregation in circular tumblers: theoretical model and scaling laws, J.
  Fluid Mech. 765 (2015) 632--652.

\bibitem{lueptow2000}
R.~M. Lueptow, A.~Akonur, T.~Shinbrot, {PIV} for granular flows, Exp. in Fluids
  28~(2) (2000) 183--186.

\bibitem{komatsu2001}
T.~S. Komatsu, S.~Inagaki, N.~Nakagawa, S.~Nasuno, Creep motion in a granular
  pile exhibiting steady surface flow, Physical Rev. Lett. 86~(9) (2001) 1757.

\bibitem{jesuthasan2006}
N.~Jesuthasan, B.~R. Baliga, S.~B. Savage, Use of particle tracking velocimetry
  for measurements of granular flows: review and application, KONA Powder and
  Particle Journal 24 (2006) 15--26.

\bibitem{eckart2003}
W.~Eckart, J.~M. N.~T. Gray, Particle image velocimetry ({PIV}) for granular
  avalanches on inclined planes, in: Dynamic Response of Granular and Porous
  Materials under Large and Catastrophic Deformations, Springer, 2003, pp.
  195--218.

\bibitem{wiederseiner2011}
S.~Wiederseiner, N.~Andreini, G.~{\'E}pely-Chauvin, G.~Moser, M.~Monnereau,
  J.~M. N.~T. Gray, C.~Ancey, Experimental investigation into segregating
  granular flows down chutes, Phys. Fluids 23~(1) (2011) 013301.

\bibitem{xiao2016}
H.~Xiao, P.~B. Umbanhowar, J.~M. Ottino, R.~M. Lueptow, Modelling density
  segregation in flowing bidisperse granular materials, Proc. R. Soc. A
  472~(2191) (2016) 20150856.

\bibitem{zhao2017}
Y.~Zhao, H.~Xiao, P.~B. Umbanhowar, R.~M. Lueptow, Simulation and modeling of
  segregating rods in quasi-2d bounded heap flow, AIChE J. 64~(5) (2018)
  1550--1563.

\bibitem{jones2018}
R.~P. Jones, A.~B. Isner, H.~Xiao, J.~M. Ottino, P.~B. Umbanhowar, R.~M.
  Lueptow, Asymmetric concentration dependence of segregation fluxes in
  granular flows, Phys. Rev. Fluids 3~(9) (2018) 094304.

\bibitem{bridgwater1980}
J.~Bridgwater, Self-diffusion coefficients in deforming powders, Powder Tech.
  25~(1) (1980) 129--131.

\bibitem{hsiau1999}
S.~S. Hsiau, Y.~M. Shieh, Fluctuations and self-diffusion of sheared granular
  material flows, J. Rheology 43~(5) (1999) 1049--1066.

\bibitem{utter2004}
B.~Utter, R.~P. Behringer, Self-diffusion in dense granular shear flows, Phys.
  Rev. E 69~(3) (2004) 031308.

\bibitem{fan2015}
Y.~Fan, P.~B. Umbanhowar, J.~M. Ottino, R.~M. Lueptow, Shear-rate-independent
  diffusion in granular flows, Phys. Rev. Lett. 115~(8) (2015) 088001.

\bibitem{fry2019b}
A.~M. Fry, V.~Vidyapati, J.~P. Hecht, P.~B. Umbanhowar, J.~M. Ottino, R.~M.
  Lueptow, Measuring segregation characteristics of industrially relevant
  granular mixtures: Part {II} -- experimental application and validation,
  submitted to Powder Tech.

\bibitem{jop2005}
P.~Jop, Y.~Forterre, O.~Pouliquen, Crucial role of sidewalls in granular
  surface flows: consequences for the rheology, J. Fluid Mech. 541 (2005)
  167--192.

\bibitem{isner2017}
A.~B. Isner, A quantitative study of size segregation in free surface granular
  flows, Ph.D. thesis, Northwestern University (2017).

\bibitem{lim2003}
S.-Y. Lim, J.~F. Davidson, R.~N. Forster, D.~J. Parker, D.~M. Scott, J.~P.~K.
  Seville, Avalanching of granular material in a horizontal slowly rotating
  cylinder: Pept studies, Powder Tech. 138~(1) (2003) 25--30.

\bibitem{zaman2016}
Z.~Zaman, An experimental study of mixing dynamics in 3d granular flows, Ph.D.
  thesis, Northwestern University (2016).

\bibitem{schlick2015c}
C.~P. Schlick, A.~B. Isner, P.~B. Umbanhowar, R.~M. Lueptow, J.~M. Ottino, On
  mixing and segregation: from fluids and maps to granular solids and
  advection--diffusion systems, Ind. Eng. Chem. Res. 54~(42) (2015)
  10465--10471.

\bibitem{schlick2016}
C.~P. Schlick, A.~B. Isner, B.~J. Freireich, Y.~Fan, P.~B. Umbanhowar, J.~M.
  Ottino, R.~M. Lueptow, A continuum approach for predicting segregation in
  flowing polydisperse granular materials, J. Fluid Mech. 797 (2016) 95--109.

\bibitem{deng2018}
Z.~Deng, P.~B. Umbanhowar, J.~M. Ottino, R.~M. Lueptow, Continuum modelling of
  segregating tridisperse granular chute flow, Proc. R. Soc. A 474~(2211)
  (2018) 20170384.

\bibitem{deng2019}
Z.~Deng, P.~B. Umbanhowar, J.~M. Ottino, R.~M. Lueptow, Modeling segregation of
  polydisperse granular materials in developing and transient free-surface
  flows, AIChE J. 65~(3) (2019) 882--893.

\bibitem{fry2019}
A.~M. Fry, P.~B. Umbanhowar, J.~M. Ottino, R.~M. Lueptow, Diffusion, mixing,
  and segregation in confined granular flows, AIChE J. 65~(3) (2019) 875--881.

\bibitem{fan2017}
Y.~Fan, K.~V. Jacob, B.~Freireich, R.~M. Lueptow, Segregation of granular
  materials in bounded heap flow: A review, Powder Tech. 312 (2017) 67--88.

\bibitem{fan2012}
Y.~Fan, Y.~Boukerkour, T.~Blanc, P.~B. Umbanhowar, J.~M. Ottino, R.~M. Lueptow,
  Stratification, segregation, and mixing of granular materials in
  quasi-two-dimensional bounded heaps, Phys. Rev. E 86~(5) (2012) 051305.

\bibitem{johanson2014}
K.~Johanson, Review of new segregation tester method by {D}r. {K}erry
  {J}ohanson, {PE}, Powder Tech. 257 (2014) 1--10.

\bibitem{lueptow2017}
R.~M. Lueptow, Z.~Deng, H.~Xiao, P.~B. Umbanhowar, Modeling segregation in
  modulated granular flow, EPJ Web Conf. 140 (2017) 03018.

\bibitem{gdrmidi2004}
G.~D.~R. MiDi, On dense granular flows, Eur. Phys. J. E 14~(4) (2004) 341--365.

\bibitem{socie2005}
B.~A. Socie, P.~B. Umbanhowar, R.~M. Lueptow, N.~Jain, J.~M. Ottino, Creeping
  motion in granular flow, Phys. Rev. E 71~(3) (2005) 031304.

\bibitem{khakhar2001}
D.~V. Khakhar, A.~V. Orpe, P.~Andres{\'e}n, J.~M. Ottino, Surface flow of
  granular materials: model and experiments in heap formation, J. Fluid Mech.
  441 (2001) 255--264.

\bibitem{isner2019}
A.~B. Isner, P.~B. Umbanhowar, J.~M. Ottino, R.~M. Lueptow, Axisymmetric
  granular flow on a bounded conical heap: Kinematics and size segregation,
  Chem. Eng. Sci. 217 (2020) 115505.

\bibitem{savageLun1988}
S.~B. Savage, C.~K.~K. Lun, Particle size segregation in inclined chute flow of
  dry cohesionless granular solids, J. Fluid Mech. 189 (1988) 311--335.

\bibitem{vanderVaart2015}
K.~van~der Vaart, P.~Gajjar, G.~Epely-Chauvin, N.~Andreini, J.~M. N.~T. Gray,
  C.~Ancey, Underlying asymmetry within particle size segregation, Phys. Rev.
  Lett. 114~(23) (2015) 238001.

\bibitem{xiao2019}
H.~Xiao, Y.~Fan, K.~V. Jacob, P.~B. Umbanhowar, M.~Kodam, J.~F. Koch, R.~M.
  Lueptow, Continuum modeling of granular segregation during hopper discharge,
  Chem. Eng. Sci. 193 (2019) 188--204.

\bibitem{tunuguntla2016}
D.~R. Tunuguntla, T.~Weinhart, A.~R. Thornton, Comparing and contrasting
  size-based particle segregation models, Comp. Part. Mech. (2016) 1--19.

\bibitem{coleman1996}
T.~F. Coleman, Y.~Li, An interior trust region approach for nonlinear
  minimization subject to bounds, SIAM Journal on optimization 6~(2) (1996)
  418--445.

\bibitem{coleman1994}
T.~F. Coleman, Y.~Li, On the convergence of interior-reflective newton methods
  for nonlinear minimization subject to bounds, Mathematical programming
  67~(1-3) (1994) 189--224.

\bibitem{taberlet2003}
N.~Taberlet, P.~Richard, A.~Valance, W.~Losert, J.~M. Pasini, J.~T. Jenkins,
  R.~Delannay, Superstable granular heap in a thin channel, Phys. Rev Lett.
  91~(26) (2003) 264301.

\bibitem{brodu2013}
N.~Brodu, P.~Richard, R.~Delannay, Shallow granular flows down flat frictional
  channels: Steady flows and longitudinal vortices, Phys. Rev. E 87~(2) (2013)
  022202.

\bibitem{baker2016}
J.~L. Baker, T.~Barker, J.~M. N.~T. Gray, A two-dimensional depth-averaged $\mu
  \mathrm{(I)}$-rheology for dense granular avalanches, J. Fluid Mech. 787
  (2016) 367--395.

\bibitem{cundallStrack1979}
P.~A. Cundall, O.~D. Strack, A discrete numerical model for granular
  assemblies, G{\'e}otechnique 29~(1) (1979) 47--65.

\bibitem{schafer1996}
J.~Sch{\"a}fer, S.~Dippel, D.~E. Wolf, Force schemes in simulations of granular
  materials, J. Phys. I 6~(1) (1996) 5--20.

\end{thebibliography}
%
%
%
\end{document}